\documentclass[]{iopart}

\usepackage{iopams}  
\usepackage{graphicx}

\usepackage{natbib}
\setcitestyle{square,super}
\newif\ifheading
\newcommand*{\fnmarkscale}{\ifheading 0.85 \else 1 \fi}
\makeatletter
\renewcommand*{\@makefnmark}
    {\hbox{\@textsuperscript{\scalebox{\fnmarkscale}{\normalfont\@thefnmark }}}}
\makeatother
\usepackage[perpage,symbol]{footmisc}

\begin{document}
\bibliographystyle{jphysicsB}

\topical[SAXS data collection and correction]{Everything SAXS: Small-angle scattering pattern collection and correction}

\author{Brian Richard Pauw}

\address{International Center for Young Scientists (ICYS)
National Institute for Materials Science (NIMS),
1-2-1 Sengen, 305-0047, Tsukuba, Japan}
\ead{brian@stack.nl}

\newpage
\tableofcontents

\begin{abstract}
%REWRITE:
For obtaining reliable nanostructural details of large amounts of sample --- and if it is applicable --- Small-Angle Scattering (SAS) is a prime technique to use. It promises to obtain bulk-scale, statistically sound information on the morphological details of the nanostructure, and has thus led to many a researcher investing their time in it over the last eight decades of development. Due to pressure both from scientists requesting more details on increasingly complex nanostructures, as well as the ever improving instrumentation leaving less margin for ambiguity, small-angle scattering methodologies have been evolving at a high pace over the last few decades. 

As the quality of any results can only be as good as the data that goes into these methodologies, the improvements in data collection and all imaginable data correction steps are reviewed here. This work is intended to provide a comprehensive overview of all data corrections, to aid the small-angle scatterer to decide which are relevant for their measurement and how these corrections are performed. Clear mathematical descriptions of the corrections are provided where feasible. Furthermore, as no quality data exists without a decent estimate of its precision, the error estimation and propagation through all these steps is provided alongside the corrections. With these data corrections, the collected small-angle scattering pattern can be made of the highest standard allowing for authoritative nanostructural characterisation through its analysis. A brief background of small-angle scattering, the instrumentation developments over the years, and pitfalls that may be encountered upon data interpretations are provided as well. 

%Lastly, good and bad examples of small-angle scattering applied in the materials science field are juxtaposed. Though this particular review scope will be particularly aimed at elucidating samples exhibiting polydisperse nanostructures, many of the details are equally applicable to other variants of scattering and other types of samples.

\end{abstract}

\maketitle

\section{Introduction}

	\subsection{Scattering to small angles}\label{sc:ssa}
	
		\begin{figure}
			\centering
			\includegraphics[angle=0, width=0.95\textwidth]{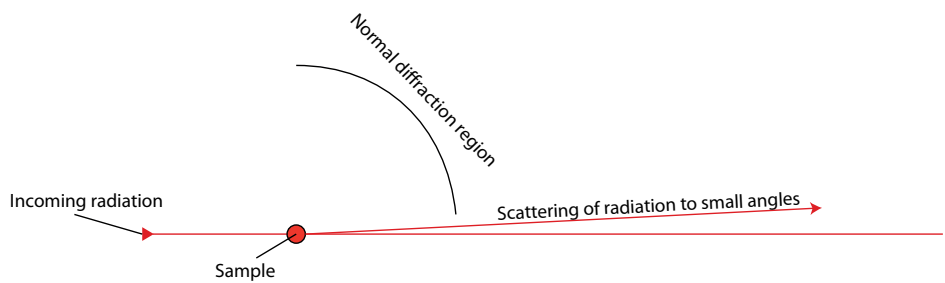} 
			\caption{The scattering of radiation to small angles by a sample (small-angle scattering). Angles normally used for diffraction analysis are also shown. \label{fg:sas}}
		\end{figure}
		The interaction of radiation with inhomogeneities in matter can cause a small deviation of the radiation from its incident direction, called small-angle scattering (Figure \ref{fg:sas}). Such small-angle scattering (SAS) occurs in all kinds of materials, be they (partially) crystalline or amorphous solids, liquids or even gases, and can take place for a wide variety of radiation, such as electrons (SAES) \cite{Schultz-1981,Carpenter-1978}, gamma rays (SAGS) \cite{Kane-1978,Kane-2005}, light (LS) \cite{Higgins-1978,Chu-2000}, x-rays (SAXS)\cite{Kratky-1933,Glatter-1982,Allen-2005} and even neutrons (SANS) \cite{Allen-2005,Belushkin-2011,Hollamby-2013}. For the purpose of this review, we shall limit ourselves to X-ray scattering, one of the more prolific sub-fields of small-angle scattering, though it should be noted that many of the principles and corrections that apply to X-rays may be applied to neutrons as well as some of the other forms. %other early examples are from 1936 corey and wyckhoff as indicated by Harget-1971
		
		The phenomenon of small-angle scattering can and has been explained in a variety of ways, with many explanations starting from the interaction between a wave and a point-shaped interacting object \cite{Glatter-1982,Feigin-1987}. For crystallographers, however, this phenomenon may be more readily understood as peak broadening of the [000] reflection (which is present for all materials), whereas for the mathematically inclined, small-angle scattering can be defined as the observation of a slice through the intensity component of the 3D Fourier transform of the electron density \cite{Debye-1949,Allen-2005,Stribeck-2007,Schmidt-Rohr-2007,Pauw-2010a}. %It is brought about when inhomogeneities are present in the material: for a homogeneous material, the incident radiation effects a homogeneous, self-cancelling waveform emission from the material. When inhomogeneities are present, this emission is no longer self-cancelling and measurable interference effects arise.
		%last sentence seems out of place...
			
		Small-angle x-ray scattering can be applied to a large variety of samples, with the majority consisting of two-phase systems \cite{Tatchev-2008}. In multiphase systems where the electron density of one phase is drastically higher than that of the remaining phases a two-phase approximation can be made \cite{Kostorz-1991}. This assumption can be done as the scattering power in SAXS is related to the electron density contrast between the phases (squared), so that the larger the difference in electron density, the larger the scattering contribution. With such a two-phase approximation, SAXS is used to study precipitation in metal alloys \cite{Fratzl-2003,Deschamps-2012}, structural defects in diamonds \cite{Shiryaev-2003}, pore structures in fibres \cite{Thuenemann-2000,Chu-2001,Pauw-2010a}, particle growth in solutions \cite{Viswanatha-2007}, coarsening of catalyst particles on membranes \cite{Smith-2008}, characterisation of catalysts \cite{Schnepp-2013}, soot growth in flames \cite{Kammler-2005}, structures in glasses \cite{Walter-1997}, void structure in ceramics \cite{Allen-2005}, and for structural correlations in liquids \cite{Temleitner-2007}, to name but a few besides the plethora of biological studies (which are well discussed in other work \cite{Koch-2003}).
		
		Small-angle scattering thus has a wide field of applicability in systems with only one or two phases. When the number of phases in the sample is increased to three, the complexity increases dramatically, drastically lowering the fields of application \cite{Tatchev-2008}. Some existing examples are studies on the extraction of hydrocarbons from coal \cite{Ciccariello-2007}, absorption studies on carbon fibres \cite{Iiyama-2004} and determination of closed vs. open pores in geopolymers \cite{Maitland-2011}. In neutron scattering, one of the three phases can sometimes be "tuned out" through smart solvent choices, essentially resulting in a scattering pattern effected by two contrasting phases. For multiphase systems straightforward SAXS is rarely attempted, though some groundwork for such applications has recently been laid \cite{Tatchev-2010}. Instead, element-specific techniques such as Anomalous SAXS (ASAXS) \cite{Walter-1997,Tatchev-2008} or combinations between SAXS and SANS \cite{Ohnuma-2009} are used to extract element-specific information.
		
		One additional drawback of SAXS, besides its preference for two-phase systems, is the ambiguity of the resulting data. As in common, straightforward SAXS measurements only the scattering intensity is collected (and not the phase of the photons), critical information is lost which prevents the full retrieval of the original structure (the ``phase problem''). As concisely explained by \citet{Shull-1947}: ``Basically it is the distribution of electron density which produces the scattering, and therefore nothing more than this distribution, if that much, can be obtained without ambiguity from the X-ray data.''. This means that a multitude of solutions may be equally valid for a particular set of collected intensities which may only be resolved by obtaining structural information from other techniques such as Transmission Electron Microscopy (TEM)\cite{Williams-2009} or Atom Probe (AP)\cite{Hono-2002,Hono-2011}. This has drastic effects on the retrievable information.
		
		In particular, of the three most-wanted morphological aspects: 1) shape, 2) polydispersity, and 3) packing, two must be known or assumed to obtain information on the third \cite{Glatter-1982,Feigin-1987,Pedersen-1997}\footnote{These three cannot be uniquely separated due to the theoretical impossibility for unambiguous separation between the interparticle- and intraparticle scattering \cite{Glatter-1982,Feigin-1987,Pedersen-1997} (i.e. it is impossible to separate shape and polydispersity from packing effects), and the impossibility to determine uniquely the particle size distribution as well as the shape of the particles from the scattering pattern \cite{Glatter-1982}. The correlation function and chord length distribution (which combine these three contributions) are however unique for a given small-angle scattering pattern.}. This can be illustrated with a few examples. By making a monodisperse assumption about the particle size distribution and assuming infinite dilution (i.e. no packing effects), the possible particle shapes become limited and can be extracted by low-resolution molecule shape resolving programs \cite{Svergun-1996}. Alternatively, knowledge on the particle size distribution and particle shape can result in a solution for the arrangement of the particles in space, as applied in structure resolving programs \cite{Toth-1992,Pusztai-1997}. Lastly, by making a low-density packing assumption and given a known particle shape (from TEM), a unique particle size distribution remains \cite{Magnani-1988,Pauw-2013,Pauw-A2013,Rosalie-A2012,Schnepp-2013}. 
		
		Despite these drawbacks, many practical applications have confirmed the validity of such small-angle scattering-derived information. For example, literature shows good agreement between TEM and SAXS analyses of gold nanoparticles \cite{Mori-2006}, krypton bubbles in copper \cite{Pedersen-1996}, commercially available silica sphere dispersions \cite{Goertz-2009}, coated silica particles \cite{De-Kruif-1988,Pedersen-1994}, zeolite precursor particles \cite{De-Moor-1999}, spherical precipitates in Ni-alloys \cite{Sequeira-1997}, and the diameter of rodlike precipitates in MgZn alloys \cite{Rosalie-A2012}, to name but a few. %show
		
\begin{figure}[bhtp]
			\begin{center}
					\includegraphics[angle=0, width=0.9\textwidth]{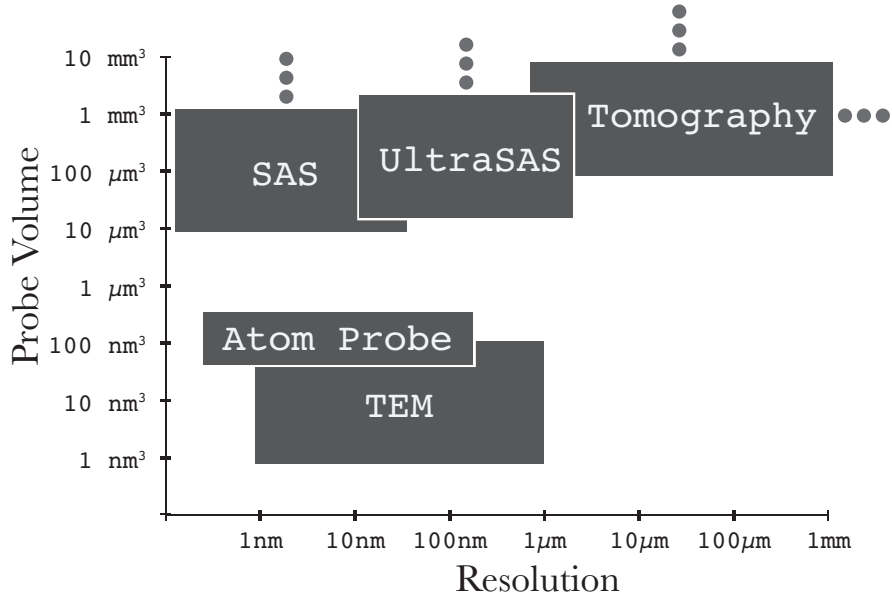}
			\caption{Typical size range of distinguishable nanostructural features (horizontal axis) and sampling volume (vertical axis) of various volumetric techniques: transmission electron microscopy (TEM), atom probe (AP), tomography and small-angle and ultra-small-angle scattering techniques (SAS/UltraSAS). Dots indicate straightforward extensibility in the indicated direction.
					\label{fg:comp}}
			\end{center}
		\end{figure}

		Small-angle X-ray scattering thus needs to be combined with supporting techniques (such as TEM, AP or porosimetry) \emph{and} is best performed on samples with two main contrasting phases. When these conditions are met, however, it will provide information on morphological features ranging from the sub-nanometer region to several micron. This information is valid for the entire irradiated volume of sample, which can be tuned from cubic micrometers to cubic millimeters and beyond (Figure \ref{fg:comp}). Furthermore, it can quantify the structural details of samples that are more challenging to quantify using electron microscopy, such as structures of glasses, fractal structures and numerous in-situ studies, as well as volume fraction and size distribution studies.
		
	\subsection{The push for better data}
	
		From the inception of SAXS around the 1930's, significant effort was expended on improving the data obtained from the instruments as it became clear to the early researchers that what you get out of it depends on what you put into it (i.e. that the quality of the results were linearly dependent on the quality of the data collected). A good overview of the early efforts is given by \citet{Bolduan-1949}. In particular, advances in collimation led to the widespread use of three collimators to reduce background scattering \cite{Bolduan-1949,Worthington-1956}, focusing and monochromatisation crystals (and even practical point-focusing monochromators \cite{Fankuchen-1937,Fankuchen-1939,Shenfil-1952,Furnas-1957}), high-intensity X-ray sources and total reflective mirrors. These early developments have led to near universal adoption of all of these elements in subsequent instruments to improve the flux and signal-to-noise ratio.
		
		X-ray sources in particular have increased drastically in brightness, leading to a similar increase in photon flux at the sample position for many small-angle scattering instruments. Where initially photon fluxes from laboratory sources were on the order of $10^3$ to $10^4$ photons per second (est. using \cite{Fankuchen-1939}). This has increased to the current flux (at the SAXS instrument sample position) from micro source tubes and rotating anode generators of about $10^7$ photons per second, useful for most common x-ray scattering experiments. For monitoring of dynamic processes, position-resolved or SAXS tomography experiments where higher flux is required, synchrotron-based instruments can deliver around $10^{11}$ to $10^{13}$ photons per second to the sample environment. The highest flux currently achievable on specialised beamlines such as BL19LXU at the SPring-8 synchrotron, fluxes of $10^{14}$ photons per second can be obtained. X-ray lasers such as SACLA in Japan, the European XFEL and the LCLS in the US provide very intense \emph{pulses} of X-rays, but the total flux is limited to about $10^{11}$ photons per second.
				
		The thus obtained increase in flux and reduction of parasitic scattering was further exploited by the advent of new detection systems. The first SAXS instruments employed step-scanning geiger counters \cite{Jellinek-1946} or photographic film (with a notable instrument even using 3 photographic films simultaneously \cite{Hermans-1959} so that sufficient information could be collected to measure in absolute units \cite{Heikens-1959}), which were rather laborious and time-consuming detection solutions. The photographic films in particular had a very nonlinear response to the incident intensity, necessitating complex corrections \cite{Chantler-1993}. The advent of image plates \cite{Cookson-1998} and 2D gas-filled wire detectors \cite{Gabriel-1982} mostly replaced the prior solutions, though image plates have a low time resolution (given the need to read and erase them), and the 2D gas-filled wire detectors suffer from a low spatial resolution due to a considerable point-spread function \cite{Le-Flanchec-1996}. Charge coupled device (CCD) detectors enjoy a modicum of success, though they suffer from reduced sensitivity alongside a slew of other issues \cite{Barna-1999}. A costly but overall relatively problem-free detector came about with the development of the direct-detection photon counting detector systems such as the linear position sensitive MYTHEN detector \cite{Schmitt-2003}, the 2D PILATUS detector \cite{Eikenberry-2003}, its upcoming successor, the EIGER detector \cite{Johnson-2012}, as well as the Medipix and PIXcel detectors \cite{Campbell-2011}. The required corrections for these detectors will be discussed in \S \ref{sc:detcor}.

 	\subsection{The next steps}
		
		A typical small-angle measurement currently consists of three steps: a rather straightforward data collection step, a data correction step to isolate the scattering signal from sample- and instrumental distortions, and an analysis step. While several works exist that detail the measurement procedure as well as the analysis \cite{Stribeck-2007}, comprehensive reviews of all possible data correction steps are less easy to find. This work therefore discusses the data collection and in particular highlights the possible data correction steps. After the data correction steps, a corrected scattering pattern of the highest of standards is obtained, which can be quite valuable. Good quality data and a good understanding of its accuracy and information content limitations greatly facilitates the process of data analysis and therefore forms the basis of any sound structural insights.
				
		%Despite the advances in instrumentation, the collected data has to be subjected to data corrections to retrieve the scattering information as cleanly as possible. 
		
		%With these advances, good data can be obtained which should be subjected to data corrections and analysis. These topics (collection, correction and analysis) will be reviewed here in detail. Data corrections will be spotlighted as they appear to be underemphasised in many current publications, and to the best of the author's knowledge, no comprehensive review covering all corrections appear to exist. 
		
		%After this, data analysis methodologies, in particular less common methodologies, will be highlighted to broaden the knowledge on the spectrum of tools available to the small-angle scatterer.

\section{Data collection}
	\subsection{The importance of good data}
	
		At the core of a good small-angle scattering methodology lies the collection of reliable, consistent data with good estimates for the data uncertainty. Once high-quality data has been collected for a particular sample, it can be forever be subjected to a variety of analyses. The data collected in the timespan of several days, during sample measurements at synchrotrons in particular, is often subjected to analysis (many) months after the measurement. Ensuring that the collected scattering pattern is an accurate representation of the actual scattering, therefore, is of the utmost importance in any small-angle scattering methodology.
		
		It almost does not need mentioning that conversely, poorly collected data should be shunned. It will confuse at best, and provide wrong conclusions at worst which could lead to disaster. Poorly collected small-angle scattering data has little to no information content in small-angle scattering, and likely consists of mostly background and parasitic scattering. In order to aid the novice researcher in collecting sufficient (and the right) information from a SAXS measurement, a data collection checklist is provided in the appendix. %which appendix?
	
	\subsection{Instrumentation}
	
		While in the past many instruments were designed and built in-house, nowadays many good instruments can be obtained from a large variety of instrument manufacturers. Given the current ease of obtaining money for a complete instrument rather than instrument development, and the drastic reduction in time required between planning and operation, the extra cost involved may in many cases be offset by the benefits. These instruments come in a variety of flavours and colours, but can essentially be divided into three main classes: 1) pinhole-collimated instruments, 2) slit-collimated instruments, and 3) Bonse-Hart instruments relying on multi-bounce crystals as angle selectors\footnote{A good review of instrumentation is also given by \citet{Chu-2001}. Furthermore, tools for instrument design evaluation have recently become available \cite{Knudsen-2013}}.
		
		\subsubsection{Pinhole-collimated instruments}
			\begin{figure}
				\centering
		   		\includegraphics[angle=0, width=0.95\textwidth]{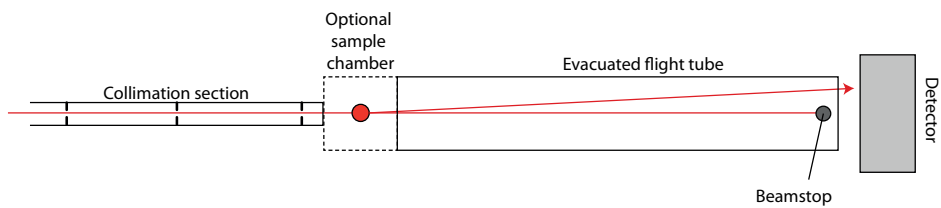} 
		   		\caption{The required components for a small-angle scattering experiment. \label{fg:sass}}
			\end{figure}

			The first of these three, pinhole-collimated instruments (schematically shown in Figure \ref{fg:sass}) have become very popular due to their flexibility in terms of samples and easy availability of data reduction and analysis procedures. While initially eschewed for slit-collimated instruments due to the drastically higher primary beam intensity of the latter, improvements in point-source X-ray generators as well as 2D focusing optics have reduced the weight of this argument somewhat. These type of instruments also dominate the small-angle scattering field at synchrotrons as well as neutron sources due to their aforementioned flexibility.
			
			%discuss source types? Wiesmann-2009 for I\muS
			These instruments typically consist of a point-based X-ray source followed by X-ray optics. These optics are either used to parallelise the photons emanating from the source, or focus the X-rays to a spot on the detector or sample. After the X-ray optics, the beam is then further collimated using either three or more collimators made from round pinholes or sets of slit blades, separated by tens of centimetres to several meter (a particular effect of the collimation on beam properties is given in \S \ref{sc:coh}). While the third collimator was required to remove slit or pinhole scattering from the second collimator \cite{Bolduan-1949,Wignall-1990,Pedersen-2004}, the recent development of single-crystalline ``scatterless'' slits remove the need for the third collimator \cite{Gehrke-1995,Li-2008}. %apparently, Hosemann already used a third pinhole in 1939 according to Wignall-1990
			
			There are two main instrument variants in circulation as to what happens after the collimation section. One type of instrument ends the in-vacuum collimation section with an x-ray transparent window, allowing for an in-air sample placement and environment before entering another x-ray transparent window delimiting the vacuum section to the detector (this sample-to-detector vacuum section is also known as the ``flight-tube''). As this introduction of two x-ray transparent windows and an air path generates a non-negligible amount of small-angle scattering background itself, it does not lend itself well to samples with low scattering power \cite{Dubuisson-1997}. The second instrument variant, therefore, consists of a vacuum chamber (and often a vacuum valve which can be closed to maintain the vacuum in the flight-tube during the sample change procedure), and thus allows an uninterrupted flightpath from collimation through the sample into the flight tube. While this generates the least unwanted scattering, it does add restrictions to the sample and sample environments that can be put in place \cite{Pedersen-2004}.
			
			At the end of the flight tube sits the in-vacuum beamstop, whose purpose is to prevent the transmitted beam from damaging the detector or causing unwanted parasitic scattering, and can be one of three types. This beamstop can be a normal beamstop, which blocks all of the transmitted beam. It is useful in many cases, however, to have an estimate for the amount of radiation flux present in the transmitted beam. For this purpose, the beamstop can be replaced or augmented with a small PIN diode, which measures the flux directly (albeit on arbitrary scale), or the beamstop can be made ``semi-transparent'', meaning that the beamstop is adapted to pass through a heavily attenuated amount of radiation which subsequently falls onto the detector. The presence of either of the two latter options can be used to benefit the accuracy of the data reduction step, leading to more accurate data and therefore more accurate results. %WHAT, NO REFERENCES?
			
			Finally, the flight tube exits in a window followed (almost) immediately by the detector. For detectors with a large detecting area, this exit window (and the flight-tube exit section) must be engineered to be strong and large, sometimes leading to visible parasitic scattering from the window material. It is therefore recommended to keep the detector small, allowing for a small and modular flight tube with very little exit window issues. Alternatively, for very modern systems, some detectors can work in-vacuum as well which removes this last (small) source of parasitic scattering and allows for step-less translation of the detector and beamstop within this vacuum, drastically increasing the flexibility in angular measurement range. 
			
			One alternative to this type of instrument was the "Huxley-Holmes" camera which contained two separate optical components for monochromatization and focusing, to achieve a very low background \cite{Zemb-2003}. While this instrument is performing well, the authors currently recommend going for a more common configuration instead consisting of focusing optics followed by scatterless slits \cite{Zemb-2013-pc}.
		
		\subsubsection{Slit collimated instruments}\label{sc:slit}
		
			A second type of instrument exists which is much more compact than the pinhole-collimated systems, is less expensive and illuminates a larger amount of sample to collect more scattering. This type of instrument is often referred to as a ``Kratky'' or ``block-collimated'' camera, perhaps best explained in \citet{Kratky-1963} and \citet{Glatter-1982}. This camera is commonly built on a line-shaped X-ray source, and collimates the x-ray beam using rectangular blocks of metal\footnote{A subsequent interesting improvement by \citet{Schnabel-1972} using glass blocks in the collimation system did not catch on, whereas beam monochromatisation and/or focusing has been a quite widely implemented improvement \cite{Fritz-2006}}. While this instrument is sometimes referred to as an ultra-small-angle scattering instrument, it is typically used as a normal small-angle scattering instrument.
			
			The line-shaped cross-section  of the X-ray beam does bring with it a major drawback, in that the collected scattering pattern is substantially different from the pattern one would obtain from a pinhole-collimated instrument. Effectively, the scattering pattern is distorted or blurred due to a superposition of intensity contributions from various scattering points along the line-shaped beam. While the collected ``slit-smeared'' scattering patterns can be subjected to a numerical correction to compensate for this smearing effect, such de-smearing processes in the best case merely amplify the noise in the system and in the worst case introduces artefacts which could be mistaken for real features \cite{Vad-2011}. This de-smearing procedure will be discussed in more detail in paragraph \ref{sc:SM}. Furthermore, analysis of samples containing an anisotropic structure becomes more tedious, leaving the instrument most suited to isotropically scattering samples.
			
			There are a number of instruments preceding the block-collimated camera, which nonetheless employed a line-shaped X-ray beam collimated with a series of slits instead \cite{Ritland-1950,Worthington-1956,Hermans-1959}. While these formed the basis of the first SAXS instruments, and are by definition slit-collimated instruments, they are no longer in widespread use.
			
		\subsubsection{Bonse-Hart instruments}\label{sc:bh}
		
			A third type of instrument is one particularly suitable for ultra small-angle scattering purposes (for the analysis of larger structures typically from several nanometer to several tens of microns), and is known as the ``Bonse-Hart'' camera \cite{Bonse-1966}. These instruments utilise the high angular selectivity of crystalline reflections to single out a very narrow band of scattering angles for observation, i.e. using the crystals as angle selectors both for collimation- as well as analysis purposes. While the idea of using crystalline reflections was not new \cite{Fankuchen-1939,Ritland-1950}, the advantage of the implementation by \citet{Bonse-1966} was the ease of use and improved angular selectivity of implementing channel-cut crystals rather than separate or single-bounce crystal elements. 
			
			The incident beam is collimated to a highly parallel beam through multiple crystalline reflections rejecting all but the angles in reflection condition. The sample is placed into this parallel beam effecting small-angle scattering as the beam passes through the sample. A second crystal (a.k.a. ``analyser crystal'') is then used to pick out a single angular band of the scattered radiation. Through rotation of the analyser crystal, the scattered intensity at various angles can be evaluated with an extremely high angular precision c.q. resolution. A few standalone instruments have been generally constructed on synchrotrons \cite{Chu-1992,Diat-1995,Ilavsky-2009a,Ilavsky-2013}, and several more have been built as complementary instruments around laboratory x-ray sources (tube- as well as rotating anode sources) \cite{Bonse-1966,Gravatt-1969,Lambard-1991,Chu-1992}.
			
			These instruments also suffer from the aforementioned smearing effect due to the essentially line-shaped incident beam, thus requiring desmearing of the data \cite{Lake-1967,Gravatt-1969}. An additional drawback to these instruments is the requirement for a step-scanning evaluation of the scattering curve (though there are efforts in neutron scattering to overcome this limitation \cite{Niel-1991}), which increases measurement times considerably. Due to the fast intensity falloff at higher angles, and the extremely narrow angular acceptance window of the analyser crystal, this instrument performs best at ultra-small angles but has much reduced efficiency at larger angles. These properties render this type of instrument a useful \emph{addition} to existing SAXS instrumentation, but is less frequently encountered as a standalone instrument. 
		
			While the difference between a Kratky camera and a Bonse-Hart camera initially seemed to be in favour of the Kratky camera \cite{Kratky-1970}, it gradually became clear that both instruments have their place in the lab. For small-angle scattering measurements on weakly scattering systems at common small angles (i.e. $0.1\leq q$ (nm$^{-1}$) $\leq 3$), a Kratky camera performs very well, while for measurements to very small angles (i.e. below $q$ (nm$^{-1}$) $\approx 0.1$) the Bonse-Hart approach would be the preferred instrument \cite{Deutsch-1980}. 
			
		\subsubsection{A note on collimation and coherence}\label{sc:coh}
					
			In typical scattering measurements, only a fraction of the volume is irradiated with \emph{coherent} radiation (i.e. with in-phase electromagnetic fields), therefore only that fraction of the irradiated sample volume contributes to the scattered intensity \cite{Veen-2004}. In other words, the irradiated sample volume typically contains a multitude of so-called ``coherence volumes'', each of which contributes to the scattering pattern. As there is no inter-volume coherence, it is the sum of the scattering intensities (as opposed to the sum of the amplitudes) from each of these volumes that is detected \cite{Livet-2007}.

\begin{figure}
			   \centering
			   \includegraphics[angle=0, width=0.65\textwidth]{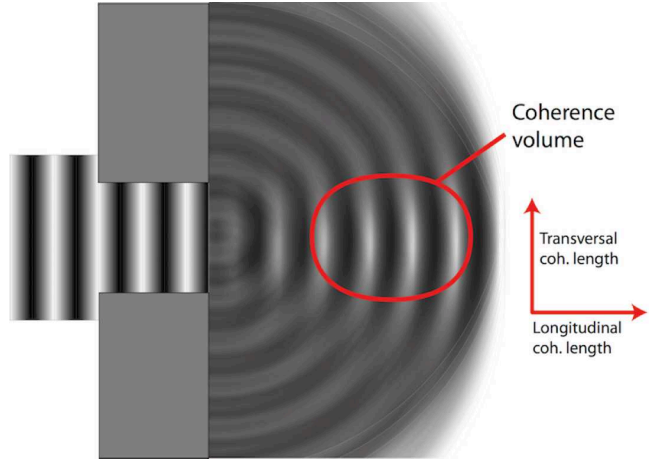} 
			   \caption{Coherence volume after a slit. The larger the slit, the smaller the transversal coherence length. \label{fg:intro_coherence}}
			\end{figure}

			These coherence volumes are defined by two components, the longitudinal component (parallel to the beam direction) and the transversal component (perpendicular to the beam direction, c.f. Figure \ref{fg:intro_coherence}). The longitudinal component is dependent on the degree of monochromaticity of the radiation, and is large for monochromatic radiation and quite small for polychromatic beams \cite{Livet-2007}. The transversal dimension $\zeta_t$ of the coherence volume is defined through the collimation, in particular through the dimensions of the beam-defining collimator and its distance to the sample, and can be estimated as \cite{Veen-2004}:
			\begin{equation}\label{eq:coh}
				\zeta_t=\frac{\lambda l}{w}
			\end{equation}
			where $\lambda$ is the wavelength of the radiation, $l$ the distance between the beam-defining collimator and the sample, and $w$ the size of the collimator opening ($\zeta_t$ can be calculated for each direction for collimators with nonuniform openings). 
			
			The estimation of the transversal coherence length is an important check for experiments. Scattering objects with dimensions close to or larger than the transversal coherence length may not contribute significantly to the small-angle scattering as the coherence volume will be within a uniform region of material (an effect seen amongst others by \citet{Rosalie-A2012}). This effect can be exploited to investigate the actual transversal coherence length in an instrument as shown by \citet{Gibaud-1996}. For a more detailed treatment of coherence (i.e. when it is approaching significance or what happens when a \emph{single} coherence volume encompasses the sample), the reader is referred to the aforementioned literature. 

\section{Data reduction and correction}

	\subsection{What corrections?}
	
		While a scattering pattern may have been recorded on the best available instrumentation, there are nevertheless some corrections to be done. The corrections must correct for as much as possible any data distortions introduced by the x-ray detection system. Further small corrections consist of spherical corrections, polarisation correction and sample self-absorption correction. More significant corrections are corrections for background, dark current or natural background, deadtime correction and scaling to absolute units. Many of these steps also need to be done in an appropriate order. These corrections will be discussed in this section, accompanied by magnitude estimates and error propagation methods where appropriate. 
		
		The goal of all these corrections is to recover and scale the collected intensity to obtain the \emph{true} scattering cross-section (which is often still called the ``intensity'' or ``absolute intensity'' colloquially) as well as an estimate for its relative $\sigma_r$ and absolute uncertainty $\sigma_a$ for all datapoints $j$ : $I_{\mathrm{true},j} \pm \sigma_{r,j} \pm \sigma_a$ (though more advanced error analysis is possible \cite{Hogg-2010}). Note that the absolute uncertainty is independent of the datapoints as it is the uncertainty estimate for the total scaling of the scattering cross-section. 
		
		It is the common consensus in the small-angle scattering community that ensuring the correct implementation of all these data corrections rests on the shoulders of the instrument manufacturer, the beam line responsible (in case of synchrotrons) or the instrument responsible. In other words, the beginning small-angle scatterer should never have to deal with these, and should receive corrected scattering cross-sections with uncertainties. The reason behind this is that in order to do most of these corrections a level of instrument understanding and characterisation is needed which cannot be expected of the casual user. In reality, however, the user can be left to their own devices and an idea on the required steps and sequence may be of some help. Several data processing packages are available to aid the user with the most pressing data correction steps \cite{Boesecke-2007,Ilavsky-2012,Knudsen-2013,Kieffer-2013} (not an exhaustive list).
		
		The purpose of this section is to introduce every possible correction, and provide a modular toolbox for constructing data correction sequences. Some corrections are "turtles all the way down", increasing in complexity the more it is investigated. For these, only the top "turtles" are given, with enough references to fine-tune the details as required. Finally, example data correction schemes are given of increasing complexity to accommodate the occasional experimentalist, the professional and the SAXS-o-philic perfectionist.
		
		%some background subtraction discussion by Stothart. It seems that there are quite a few papers discussing only one aspect of data correction, hardly any comprehensive papers. Strunz also has a nice correction, as does br\UTF{00FB}let
		
	\subsection{Data reduction steps and sequence}
	
		The required data steps ordered by their approximate position in the data reduction and correction sequence is indicated in Table \ref{tb:corr}. Where applicable, the paragraph in which the data correction in question is discussed is provided. Convenient two-letter abbreviations have also been provided. While the table includes a fair few corrections and is suitable to a variety of detectors, it should not be considered universal as some detectors are in need of more corrections, or application of the corrections in a slightly different order. 
	
\begin{table}[htdp]
\begin{footnotesize}
\begin{center}
\caption{The data reduction and correction steps in approximate order of application. The abbreviations in the ``Abbrv.''-column is listed in the appendix. The columns ``$\sigma_r$'' and ``$\sigma_a$'' indicate whether a correction affects either the relative uncertainty and/or the absolute uncertainty (if so, column marked with $\circ$). Also indicated are four types of detectors (CCD: a typical CCD detector with tapered fibres or image intensifier, IP: image plate, DD: Direct detection systems such as the hybrid pixel detectors, and WD: Wire detectors and similar), whose columns indicate the severity of the effect of each correction for that detector, where ``+'' indicates the correction has to be applied by manufacturer or user, ``-'' indicates a minor correction that can be ignored. Complexity (Cx) column indicates the approximate complexity of the correction implementation, with 0 being easy, and 3 complicated.\label{tb:corr}}%, the ``apply to''-column indicates what data the correction should apply to, with: s=Sample measurement, sb=Sample background measurement, nb=natural background/darkcurrent measurement .}

\begin{tabular}{ p{0.03\textwidth} p{0.05\textwidth} p{0.35\textwidth} p{0.04\textwidth} p{0.01\textwidth} p{0.01\textwidth} p{0.04\textwidth} p{0.01\textwidth} p{0.015\textwidth} p{0.015\textwidth} p{0.04\textwidth}}
		\hline
		Step no. & Abbrv. & Description & \S & $\sigma_r$ & $\sigma_a$ & CCD & IP & DD & WD & Cx\\
		\hline
		1. & DS & Data read-in corrections for manufacturer's data storage peculiarities &\ref{sc:DS} && &+&+&+&+ &0-3\\
		2. & DZ & Dezingering - removing high-energy radiation streaks & \ref{sc:DZ} && &+&-&-&-&2\\
		3. & FF & Detector flat-field correction & \ref{sc:FF} & $\circ$ & &+&-&+&+&1\\
		4. & DT & Detector dead-time correction (photon counting detectors) & \ref{sc:DT} & $\circ$ & &-&-&-&+&2\\
		5. & GA & Detector non-linear response (gamma-)correction &\ref{sc:GA} &$\circ$& &+&-&-&+&1\\
		6. & TI & Normalise by measurement time &\ref{sc:TI} &$\circ$&$\circ$ &+&+&+&+&0\\
		7. & DC & Subtraction of natural background or dark current measurement (itself subjected, when applicable, to steps 1-6) & \ref{sc:DC} && &+&+&+&+&0\\

		8. & FL & Normalize by incident flux &\ref{sc:FL} &$\circ$&$\circ$ &+&+&+&+&0\\
		9. & TR & Normalize by transmission &\ref{sc:TR}&$\circ$&$\circ$ &+&+&+&+&0\\

		10. & GD & Detector geometric distortion correction & \ref{sc:GD} &$\circ$& &+&$\pm$&-&+&3\\
		11. & SP & Spherical distortion correction (area dilation)& \ref{sc:SP} &$\circ$& &-&-&-&-&1\\
		12. & PO & Correct for polarisation (even for unpolarised beams) & \ref{sc:PO} &$\circ$& &-&-&-&-&1\\
		13. & SA & Correct for sample self-absorption & \ref{sc:SA} &$\circ$& &-&-&-&-&1-3\\

		14. & BG & Subtract background (itself subjected to steps 1-11)  & \ref{sc:BG}&$\circ$&$\circ$ &+&+&+&+&0\\
		15 & TH & Normalise by sample thickness & \ref{sc:TH}&&$\circ$ &+&+&+&+&0\\
		16. & AU & Scale to absolute units &\ref{sc:AU} &&$\circ$ &+&+&+&+&1\\
		
		17. & MK & Mask dead and/or shadowed pixels & \ref{sc:MK} && &+&+&+&+&0\\				
		18. & MS & Correct for multiple scattering* & \ref{sc:MS} &$\circ$& &-&-&-&-&3\\
		19. & SM & Correct for beam shape smearing effects* & \ref{sc:SM} &$\circ$& &-&-&-&-&3\\
		20. & -- & Radial or azimuthal averaging & \ref{sc:int} &$\circ$& &&&&&0\\
		\hline
		\multicolumn {10}{p{0.9\textwidth}}{*) These are more robustly dealt with by smearing the data fitting model rather than desmearing the data}\\
		\hline

\end{tabular}
\end{center}
%\label{default}
\end{footnotesize}
\end{table}%

	\subsection{Detector corrections: DS, DZ, FF, DT, GA, DC, GD, MK }\label{sc:detcor}
		
		In order to detect x-rays, a wide variety of detectors have become available. Depending on the detection method, imperfections and physical limitations may cause a deviation of the detected signal from the true signal (the number of scattered photons). In a perfect case, you would measure the same (true) scattering signal irrespective of the type of detector used. 
		
		Real detectors, however, have imperfections, tradeoffs and drawbacks. Some of these detectors and their individual drawbacks will be discussed here, after elaboration on the possible distortions. The distortions can generally be divided into two categories, intensity distortions and geometry distortions. Intensity distortions are deviations in the \emph{amount} of measured intensity, and geometry distortions are deviations in the \emph{location} of the detected intensity. First and foremost, there are data read-in corrections to consider.

		\subsubsection{Data read-in corrections: DS }\label{sc:DS}
		
			The first step for any data correction is to read in the information from detectors. While for point- and linear position sensitive detectors (PSDs), the choice has almost universally been made for the convenience of ASCII data, for image detectors this has not been so straightforward.
			
			Therefore, whenever a detector system is bought, particular attention needs to be paid to the data format of the images one obtains. For some reason, quite a few detector manufacturers worldwide prefer their own image data formats over more standard image formats (a list of some of these formats can be found in the documentation accompanying the NIKA package \cite{Ilavsky-2012}). This tendency hinders data preservation efforts (though one should preserve corrected and reduced data rather than the original data, a point discussed in \S \ref{sc:universalstorage}) and sometimes causes read-in issues of the data in data reduction packages. Two cases in particular have come to the attention of the author, the Rigaku data format and the Bruker data format, which will be used to illustrate the issue.
	
			The Rigaku data format has all the characteristics of a 16-bit TIFF image, and will actually load as such. Without going into details, 16 bits (per image value) would get you a maximum per-pixel value of 2$^{16}$: 65536. This value would be insufficient for storing the number of photons obtained for example from the aforementioned PILATUS hybrid pixel detector, which therefore uses a 32-bit image format. The Rigaku format treats such count numbers slightly differently in order to store them in 16 bits: 15 bits behave like normal bits up to a value of 2$^15$ (32768), the 16th bit acts not as a standard bit but as a ``multiply-by-32''-flag\footnote{Not quite true, the 16th bit acts as a "multiply-by" flag, with the actual integer listed in the image header}. While this is documented \cite{Rigaku-2013}, the danger lies in the compatibility of their data format with standard binary data: the intensities will be wrong, but the scientist ignorant of this issue will not immediately notice something is awry. 
			
			The Bruker data format, on the other hand, is unlikely to be compatible with any standard image reading routines, and authoritative information on the image format is not very easy to obtain. Some of their image formats appears to use an 8-bit image format (i.e. with per-pixel maximum values of 256), with a subsequent "overflow" list detailing pixels that have exceeded this 8-bit limit. Implementation and read-in of this data is therefore cumbersome, perhaps even unnecessarily complicated given the alternatives.
			
			In the best case, detector systems adhere to known and common image formats \cite{Eikenberry-2003}. Active development is ongoing for supporting detector data of these and more complicated multi-chip detectors and instruments in the NeXus format \cite{Klosowski-1997,Konnecke-2006}. The NeXus format itself is based on the very versatile, portable, well-documented and open HDF5 data storage format \cite{Folk-2010}. Such standards will hopefully resolve some of the challenges related to data ingestion into data reduction procedures.
				
		\subsubsection{Dezingering: DZ }\label{sc:DZ}
		
		Spurious signals can be detected for a range of reasons: from external sources such as cosmic rays, nearby x-ray sources or atmospheric radioactive decay, or from internal sources such as the employed electronics. These often appear as spikes or streaks in the detected signal, varying in location and among from image to image. Integrating (e.g. CCD) detectors without energy discrimination are most heavily affected by these phenomena, whilst photon-counting, energy discriminating detectors often only show a single extra count (or streak of 1 extra count) upon event occurrence. 
		
		Given their potentially high values, zingers can significantly affect the recorded signal, and should be removed in CCD-based detectors. The trick for their detection and subsequent removal is to record multiple images per measurement and mask all statistically significant differences. A suitable computational procedure is described by \citet{Barna-1999} and \citet{Nielsen-2009}.
		
		\subsubsection{Flatfield correction: FF }\label{sc:FF}
		
			Every detector apart from point detectors (i.e. every spatially resolved detector) has to be corrected for interpixel sensitivity, with the notable exception of image plates\footnote{Image plates (due to their positioning uncertainties during read-out) cannot be corrected for this effect and it is fortunate that it appears to play a very minor role in its accuracy \cite{Ito-1991}.}. As no two detection surfaces (pixels) are exactly the same due to manufacturing tolerances, slight damage or differences in the underlying electronics, to name but a few. These interpixel sensitivity variations can easily be on the order of 15\% for some detectors \cite{Tate-1997}. The correction is straightforward in theory: collect a uniform, high amount of scattering on the detector, assume the per-pixel detector response should be identical for this scattering, and use the relative difference in detected signal between the pixels as a normalisation matrix for future measurements. In practice, though, uniformly distributing a large number of photons (of the right energy) on the detector surface can be a challenge. 
			
			One solution is to irradiate directly with a low-power x-ray source placed some distance from the detector, as discussed in detail by \citet{Barna-1999}. This solution needs small corrections for area dilation and air absorption, in addition to a few more detector-specific ones, and needs a separate check of the uniformity of the source. The advantage is that it can be tuned to the energy of interest, and that a sufficient number of photons is easily acquired \cite{Procz-2011,Dinapoli-2011,Ghosh-1999}.
			%get Moy-1996 from Barna-1999 ref 7 
			
			Alternatively, doped glasses can be used to obtain a flat-field image, as suggested by \citet{Moy-1996}. This has the advantage of reduced complexity in setting up the flat-field measurement, but may suffer from nonuniform images \cite{Barna-1999} and has a reduced photon flux. One solution is to use the uniform scattering of water as flat-field measurement data, despite water not scattering uniformly at very small angles (though this can be corrected for), and the scattering intensity at larger angles being quite low for obtaining good per-pixel statistics \cite{Narayanan-2001}. Similarly, samples with known scattering behaviour can be used for such purposes \cite{Ghosh-1999}. Another solution common in laboratory settings is the use of radioactive sources (emitters) which can be easily accommodated in most instruments \cite{Ne-1993}. The major drawback of that solution is the differences between the emitter energy and the energy used during normal measurements, and a very low detected count rate necessitating impractically long collection times for decent flat-field images. The alternative suggested by \citet{Ne-1993} is the image collection during slow and well-controlled scanning of an emitter over the detector surface, with the challenge of achieving a homogeneous exposure. 
			
			The alternatives which place the radiation source at the sample location share one further advantage in case of detectors using phosphorescent screens. The advantage is that through placement of the radiating source at the sample location, one simultaneously corrects for the dependency of the response of phosphorescent screens to the direction of incident photons. If this is not done, one might consider correcting for this effect separately \cite{Barna-1999}. %danger of correcting for solid angle/area dilation at the same time!
			
			Given these challenges, it is therefore recommended for (time-stable) detectors to obtain flat-field images from the manufacturer who should be equipped to record these. The corrected intensity $I_{j\mathrm{cor}}$ for datapoint $j$ can be retrieved from the input intensity $I_j$ using a flatfield image $F_j$ (which can be normalised to 1 to avoid large numbers):
			\begin{equation}\label{eq:FF:I}
				I_{j,\mathrm{cor}}=\frac{I_j}{F_j}
			\end{equation}		
			If there are uncertainties available when performing this step, they will propagate as:
			\begin{equation}\label{eq:FF:sr}
				\sigma_{r,j,\mathrm{cor}}= I_{j,\mathrm{cor}} \sqrt{ \left[ \frac{\sigma_{r,j}}{I_{j}} \right]^2+\left[\frac{\sigma(F_j)}{F_j}\right]^2}
			\end{equation}					
			
		\subsubsection{Deadtime correction: DT }\label{sc:DT}
		
			After arrival of a photon on a detection surface or in a detection volume, a certain amount of time is needed for the detector to recover from this event before a second photon can be detected. This time is called the ``dead time'': a photon arriving in this timespan will not be detected. More precisely, the electronic pulses generated by the arrival of two near-simultaneous photons will start to overlap, causing either rejection of both photons due to the compound pulse being too high (energy rejection), or the two pulses being counted as one. This is discussed clearly by \citet{Laundy-2003}. 
			
			This correction can be unnecessary for some of the modern hybrid detectors at the count-rates they are commonly subjected to. The PILATUS detector, for example, shows a $>2$\% deviation from a linear response at an incident photon rate of more than 450000 photons per pixel per second \cite{Kraft-2009}. Gas-based detectors, especially 1D and 2D wire detectors very much need this correction. 
			
			One aspect of this correction that is of high importance is that when the data uncertainty is calculated based on counting statistics (i.e. Poisson statistics), these uncertainties should be calculated from the \emph{detected} photons, not from the deadtime-corrected photons. This implies that there is a count rate characteristic for each detector beyond which the data accuracy decreases! This phenomenon is evident from \citet{Laundy-2003}. 
			
			The number of deadtime-corrected counts $I_{j\mathrm{cor}}$ can be obtained from the detected number of counts $I_j$ collected in time $t$ by numerically finding a solution for \cite{Laundy-2003}:
			\begin{equation}\label{eq:DT:I}
				I_j=I_{j,\mathrm{cor}} \exp (-I_{j,\mathrm{cor}} T)
			\end{equation}
			with 
			\begin{equation}\label{eq:DT:T}
				T=\frac{\tau_1+\tau_2}{t}
			\end{equation}
			where $\tau_1$ is the minimum time difference required between a prior pulse and the current pulse for the current pulse to be recorded correctly. Similarly, $\tau_2$ is the minimum arrival time difference required between the current pulse and a subsequent pulse for the current pulse to be recorded correctly. As pulses follow an asymmetric profile like a log-normal function, these two times can be different (for a 1\,$\mu$s pulse shaping time this can be $\tau_1=3.0$\,$\mu$s and $\tau_2=2.0$\,$\mu$s \cite{Laundy-2003}).
			
			At this point we can also estimate the uncertainty (standard deviation) $\sigma_{r,j}$ for the corrected counts through \cite{Laundy-2003}:
			\begin{equation}\label{eq:DT:sr}
				\sigma_{r,j}=\left[ \frac{ \left(1- I_{j,\mathrm{cor}}  T \right)^2 I_j}{ 1+ 2 \exp \left(- I_{j,\mathrm{cor}} \frac{\max(\tau_1,\tau_2)}{t} \right) - 2 \left( 1+ I_{j,\mathrm{cor}}  T \right)\exp \left(- I_{j,\mathrm{cor}} T \right) } \right]^{\frac{1}{2}}
			\end{equation} 
			Interestingly, if $\tau_1$ and $\tau_2$ are known, the true uncertainty $\sigma_{r,j}$ can be retrieved from the deadtime corrected values through insertion of eq. \ref{eq:DT:I} into eq. \ref{eq:DT:sr}, which may be of use in detector systems where the deadtime correction is performed by the detector system itself.
			
		\subsubsection{Gamma correction: GA }\label{sc:GA}
		
			Most non-photon counting detectors do not necessarily give an output linearly proportional to the incident amount of radiation. This used to be especially severe for films, which required accurate corrections for each film type \cite{Chantler-1993}. For more modern detection systems the effect appears small (i.e. on the order of 1\%), but may nevertheless be considered especially when approaching the limits of the dynamic range \cite{Naday-1994,Naday-1994a,Hammersley-1995}. It is relevant for image plates \cite{Miyahara-1986,Cookson-1998,Ne-1993,Barnea-2011} and may be considered for some CCD detectors as well \cite{Hammersley-1995}. It may even be relevant for some gas-based photon-counting detectors insofar it is not already accounted for with the deadtime correction \cite{Barnea-2011}. 
			
			This correction can be applied by characterising the detector response for various \emph{fluxes} of incident radiation, for example through attenuating monochromatic radiation using a series of calibrated foils to reduce the incident flux  \cite{Hammersley-1995}. Simply collecting radiation for a longer time may obfuscate the detector response to incident flux with other time-dependent effects especially for image plates \cite{Cookson-1998}, unless this is explicitly taken into account \cite{Hammersley-1995}. Furthermore, the energy of the incident radiation has to be identical to the energy used for normal measurements, as the gamma correction can be energy dependent \cite{Ito-1991}. 
					
			One alternative solution to circumvent the need for this correction is to determine the range of incident radiation amounts where the detector response is linear, and to stay within that range. For samples which exhibit scattering covering a wider dynamic range than thus supported, attenuators can be devised in the beam path to locally attenuate the signal \cite{Narayanan-2001}. Introducing additional elements into the beam path may, however, cause scattering or act as a high-pass energy filter leading to ``radiation hardening'', and such modifications should therefore not be applied without thorough considerations of the consequences. 
	
			Lastly, while it cannot be considered a true nonlinearity correction, for image plates the measured intensity is also a function of measurement time (i.e. the delay after exposure before measuring) \cite{Ne-1993}. Internal decay causes a reduction of the measurable signal over time, with a fast decay component (with a half-time on the order of minutes) and a slow decay component (on the order of hours). Effectively, this can even cause intensity variations on the order of several percent during the read-out of the image plate. A decay time correction should therefore be considered for accurate reproduction of intensity, and such a correction is described amongst others by \citet{Hammersley-1995}. It should be noted that this time decay is likely also dependent on the energy of the used x-rays as it is for protons \cite{Bonnet-2013}.
			
			This correction is applied if the nonlinear behaviour of the intensity can be expressed as a function of the incident radiation $\gamma(I)$:
			\begin{equation}\label{eq:GA:I}
				I_{j,\mathrm{cor}}=\gamma(I_j)
			\end{equation}
			The relative datapoint uncertainty scales similarly:
			\begin{equation}\label{eq:GA:sr}
				\sigma_{r,j,\mathrm{cor}}=\sigma_{r,j}\frac{\gamma(I_j)}{I_j}
			\end{equation}
			
			%percent value also hammersley
			%check out Chu-2001 section on detectors.
			
		\subsubsection{Darkcurrent and natural background correction: DC }\label{sc:DC}
		
			There are two factors adding to the detected signal even without the presence of an x-ray beam, these are the detector ``dark current'' and the omnipresent natural radiation. While these are two separate effects, their correction is identical and can be simultaneously considered. The cause of the dark current signal depends on the detector: Some detector electronics add their own ``pedestal'' bias to prevent negative voltages entering the analog-to-digital converter (ADC) \cite{Barna-1999,Narayanan-2001}, which can be considered a form of dark current. CCD chips may also exhibit a baseline noise "read noise", photomultiplier tubes (PMTs) in image plate systems detect a small leak current without any incident photons and ion chambers also detect a small current without radiation. Natural background radiation furthermore adds a constant level of noise in any detector \cite{Ne-1993}. 
			
			The dark current components are homogeneously distributed over the entire detector, and can thus (for statistical purposes) be corrected for by subtraction of a single value from each detected pixel value. This single value is a summation of all three dark current components: a time-independent component, a time-dependent component and a flux-dependent component. To elaborate, the time-independent component would be the base amount (``pedestal'')-level, applicable to detectors based on PMTs and CCDs \cite{Stasio-2006}. Naturally occurring background radiation can be considered part of the time-dependent component, visible in every detector. One important note here is that the image plates start collecting natural radiation from the time of their last erasure rather than from the start of the measurement \cite{Ne-1993}. Some detectors may also show a time-dependent dark current in addition to the natural background \cite{Naday-1994a}. These two components can be easily determined through evaluation of the total detected signal as a function of exposure time without an applied X-ray beam. The last component, the flux-dependent dark current level is a specific complication encountered in some image-intensifier-based CCD detectors, and requires the simultaneous determination of the dark current signal alongside the measured signal through partial masking of the detection surface with x-ray absorbent material \cite{Pontoni-2002,Narayanan-2001}.
			
			This can be expressed mathematically as:
			\begin{equation}\label{eq:DC:I}
				I_{j,\mathrm{cor}}=I_j-(D_a+D_b t+D_c(\int_j I_j))
			\end{equation}
			where $D_a$ is the time-independent component, $D_b$ the time-dependent factor times the measurement time $t$, and $D_c$ the flux-dependent component for those detectors suffering from that particular complication (determined simultaneously with the measurement). Image plates furthermore have a natural decay which means that the time-dependent component may not be truly linear over large timescales. It is therefore best practice to determine the dark current contribution using exposure times similar to the measurement times. For accurate determination of the dark current contribution when measurement times are small, the averaging of multiple exposures on the time-scale of the measurement can improve statistics \cite{Narayanan-2001}.
			
			As the dark current is ideally pixel-independent, $D_a$, $D_b$ and $D_c$ can be determined to high precision when averaged over the entire detector. This should render their relative uncertainties $\sigma(D)/D$ rather small thus only having a minor effect on the intensity uncertainty. The uncertainty should propagate (assuming Poisson statistics) approximately as:
			\begin{equation}\label{eq:DC:sr}
				\sigma_{r,j,\mathrm{cor}}=\sqrt{ \sigma_{r,j}^2+\sigma(D_a)^2+ (t \sigma(D_b))^2 + \sigma\left(D_c(\int_j I_j)\right)^2 }
			\end{equation}
					
		\subsubsection{Geometric distortion: GD }\label{sc:GD}
			
			Among the more complicated detector corrections is that of the geometric distortion, which can be severe for some detectors (in particular for wire detectors and image-intensifier-based CCD detectors), small for others (i.e. $<$1\% for fibre-optically coupled CCD detectors) \cite{Naday-1994a}, to non-existent for direct-detection systems. The electronics and design of image intensifiers in CCD cameras and electronics of wire-detectors can give rise to pixels being assigned incorrect geometric positions, leading to geometric distortion \cite{Barna-1999}. Even image plate readers can show this effect due to the read-out mechanics \cite{Le-Flanchec-1996}, and it therefore seems a necessary correction for all detectors save those dependent on direct-detection (e.g. the PILATUS detector) In order to put the detected pixels back in their right "place", i.e. in a location corresponding to the arrival location of the detected photon on the detector surface, a geometric distortion correction must take place. 
			
			The most common method for this is to place a mask with regularly spaced holes in front of the detector, which is subsequently irradiated with more-or-less uniform photons originating from the sample position. This then allows for the evaluation of where on the detector the photons are observed versus where the photons actually arrived through the holes in the mask \cite{Le-Flanchec-1996,Thomas-1989}.  
			
			These corrections only really can take care of smoothly varying distortions, and are ill-suited for corrections of abrupt distortions as those found upon occurrence of discontinuous shears in fibre-optically coupled detectors \cite{Barna-1999}. Corrections for these distortions must be considered separately \cite{Deckman-1986}. Rather than trying to correct the actual image by e.g. inserting or interpolating pixel values (e.g. \cite{Thomas-1989}), one good way of dealing with these corrections is to determine a coordinate lookup table (``displacement maps'') for each pixel. These maps can subsequently be used during the data averaging procedure (c.f. \S \ref{sc:int}) to put the detected datapoints in the right bins \cite{Kieffer-2013-pc,Kieffer-2013}.
			
			Image plates, besides the small geometric distortion mentioned above also require a specific correction: one that corrects for variance in subsequent placements of image plates. Since it is mechanically challenging to reproducibly place an image plate to within 50 micron (or approximate grain size), every image plate may be slightly offset. The variance in placement for a given image plate placement and read-out procedure (ideally designed to minimise placement variance) can be quantified and evaluated for significance of severity. If necessary, symmetry in the scattering patterns can be exploited to determine the beam centre of every image. A procedure for achieving this is described by \citet{Le-Flanchec-1996}. 
			
			Due to the detector specificity of the required correction and the relatively complex procedure, the methods for correcting image distortions are not reproduced here. Geometric distortions should not affect the datapoint uncertainties.
			
			%le flanchec talks about the pixel-filling method by Castleman-1979, specifically do not use Schlien-1979 for its complexity. Both not available, first is a book, second is not digitised.

			%Barna has a good summary
			%check ref 2, 6, 7, 8 in Hammersley-1995
		%FROM HERE
		\subsubsection{Masking of incorrect pixels: MK }\label{sc:MK}
			
			In virtually any detection system there will be ``broken'' pixels, either pinned to the maximum or minimum value, or simply giving incorrect response to the incident radiation. Additionally, pixels masked by the beamstop or the beamstop holder should be ignored as well. For masking these, an oft used technique is to record a scattering pattern of a strong scatterer, after which a boolean array can be manually generated, indicating for each pixel whether it should be masked or not. For space-saving purposes, this boolean array can be reduced to a list of pixel indices to be masked.
			
			This array (or list of pixel indices) can subsequently be used in the averaging procedure to not consider invalid pixels in the procedure. Such masking does not affect the uncertainties.
		
	\subsection{Other corrections}
	
	There are a range of corrections to be done that are independent from the type of detector used. These are corrections for e.g. sample transmission (closely related to the background subtraction), correction for polarisation and area dilation. Included in these correction is the correction (or rather the scaling) to go from ``intensity'' to scattering cross-section which can later be used to retrieve volume fractions or number of scatterers to a reasonably good degree (with an expected accuracy $\sigma_a/I$ of about 10\%). 
	
		\subsubsection{Polarization correction: PO }\label{sc:PO}
		
			The scattering effect of photons depends on the polarisation of the incident radiation and the direction of the scattered radiation \cite{Soper-2011}. This phenomenon causes a slight reduction in intensity. While this effect is commonly corrected for in wide angle diffraction studies, it is often considered negligible in small-angle scattering data correction \cite{Pedersen-1991,Boesecke-1997,Portale-2007,Nielsen-2009}. When quantified, the correction amounts to nearly 1\% for scattering angles $2\theta$ of about 5 degrees. This correction applies both to unpolarised radiation as well as polarised radiation, in the former the correction is isotropic, and in the latter anisotropic. Depending on the angular range collected, the polarisation correction may be considered for a slight increase in accuracy.  
		
		The correction factor for 2D detector images is given by \citet{Hura-2000} as:
		\begin{equation}\label{eq:PO:I}
			\begin{array}{rl}
			I_{j,\mathrm{cor}}=&I_j \left[ P_i \left(1-\left(\sin(\psi)\sin(2\theta)\right)^2\right) + \right. \\
			 & \left. (1-P_i) \left(1-\left(\cos(\psi)\sin(2\theta)\right)^2\right) \right] \\
			 \end{array}
		\end{equation} 
		
		Where $\psi$ is the azimuthal angle on the detector surface (defined here clockwise, 0 at 12 o'clock) $2\theta$ the scattering angle, and $P_i$ the fraction of incident radiation polarised in the horizontal plane (azimuthal angle of 90 degrees)\footnote{A 2D solution more tuned to crystallographic studies is given by \citet{Azaroff-1955}.}. The correction for unpolarised radiation is achieved when $P_i=0.5$, most synchrotron beam lines have a $P_i\approx0.95$. As this is a correction between datapoint values, only the relative uncertainty $\sigma_{r,j}$ is affected similarly to the effect of polarisation on the intensity:
		\begin{equation}\label{eq:PO:sr}
			\begin{array}{rl}
			\sigma_{r,j,\mathrm{cor}}=&\sigma_{r,j}\left[ P_i \left(1-\left(\sin(\psi)\sin(2\theta)\right)^2\right) + \right. \\
			& \left. (1-P_i) \left(1-\left(\cos(\psi)\sin(2\theta)\right)^2\right) \right]
			\end{array}
		\end{equation}
		
		\begin{figure}
			\centering
   			\includegraphics[angle=0, width=0.95\textwidth]{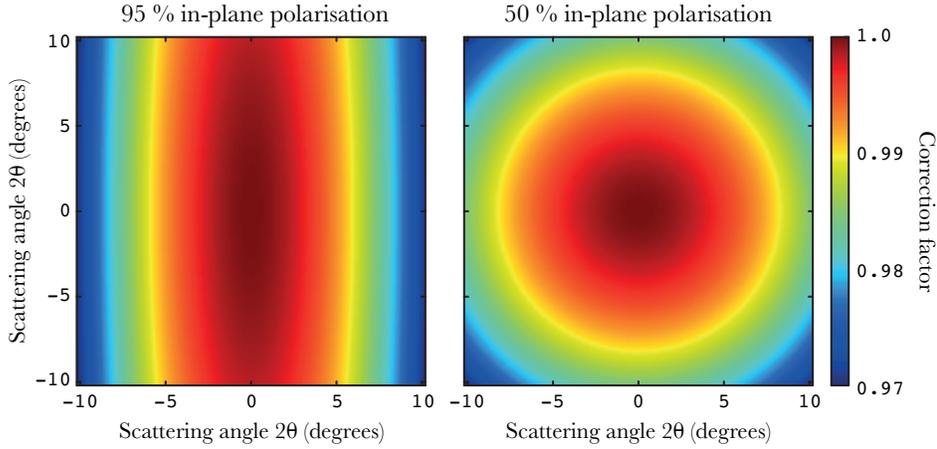} 
   			\caption{Correction factor for two-dimensional detectors given 95\% in-plane polarisation (typical reported value for synchrotrons), and no polarisation (c.q. 50\% in-plane polarisation)}\label{fg:PO}
		\end{figure}
		%check polarisation with Bendert-2013?

		\subsubsection{Transmission and flux corrections: TR and FL }\label{sc:TR}\label{sc:FL}
			 
			Any material inserted into the beam absorbs a certain amount of radiation \cite{Hura-2000,Arai-2005}. This affects the amount of background scattering impinging on the detector as well as the amount of scattering of the remaining radiation by the sample (differing slightly depending on the path through the sample as well as shown in \S \ref{sc:SA}). As the amount of radiation scattered by the sample is typically small, the absorption c.q. transmission factor can be determined by measuring the flux directly before and after the sample. 
		
		There are three commonly applied methods for measuring the sample absorption, one ``in-situ'' method commonly found at synchrotrons and two offline methods. At synchrotrons, so-called ionisation chamber detectors can be installed (usually in air) directly before and after the sample position. These detectors are very straightforward in their construction, typically consisting of two electrodes suspended in air \cite{Seemann-1949,Tanaka-1959,Miller-2013}. They are particularly suitable as they exhibit no parasitic scattering and can be used to monitor the incident flux as well as the absorption over the duration of the measurement. Assuming a non-identical but linear response for both upstream $u$ and downstream $d$ detectors, the readings without sample (indicated with subscript $_0$) and after sample insertion (subscripted $_1$) can be used to calculate the transmission factor $T_r$ through: 
		\begin{equation}\label{eq:ictransmission}
			T_r=\frac{\frac{I_{d,1}}{I_{u,1}}}{\frac{I_{d,0}}{I_{u,0}}}=\frac{ I_{d,1} I_{u,0}}{ I_{u,1} I_{d,0} }
		\end{equation}	

		It is not always possible to insert ionisation chambers, for example when working with a completely evacuated instrument. In that case, two alternative solutions can be applied to measure the beam flux sequentially before and after insertion of the sample. In one solution, the beamstop is modified to either: 1) allow for a small fraction of the direct beam to pass through and be detected by the main detector (i.e. a ``semitransparent beamstop''), or 2) where the beamstop is augmented with a small\footnote{These can even be made very small for microbeam applications as shown by \citet{Englich-2011}} pin-diode measuring the direct beam flux \cite{Morita-2007,Ellis-2003}. The second option is to place a strong scatterer in the beam downstream from the sample position, and measure the integrated scattering signal from the strong scatterer \cite{Pedersen-2004}. For the beamstop modification case, the ratio of the two fluxes (before and after insertion of the sample) is the transmission factor. In the last case, the ratio of the two \emph{integrated} intensities on the detector is the transmission factor:
		\begin{equation}\label{eq:simpletransmission}
			T_r=\frac{I_1}{I_0}
		\end{equation}
		where $I_0$ is the intensity of the primary beam without sample, and $I_1$ the intensity of the beam after insertion (and downstream) of the sample. A transmission factor correction for highly absorbing samples scattering to wide angles is discussed in \S \ref{sc:SA}. The transmission factor is dependent on the linear absorption coefficient $\mu$ and thickness $d$ of a material through:
		\begin{equation}\label{eq:TR:T}
			T_r=\exp\left(-\mu d \right)
		\end{equation}
		
		The transmission correction can be applied by dividing the detected intensity with the transmission factor. Furthermore, the detected intensity is proportional to the incident flux on the material $f_s$, which can be similarly corrected for:
		\begin{equation}\label{eq:TR:I}
			I_{j,\mathrm{cor}}= \frac{I_j}{T_r f_s} 
		\end{equation}
				
		The relative intensity uncertainty remains largely unaffected by this correction (if the background is small and/or shows little localisation), but the absolute uncertainty is directly related to the uncertainties of the measured transmission and flux:
		\begin{equation}\label{eq:TR:sa}
			\sigma_{a,\mathrm{cor}}=  \sum\limits_j I_{j,\mathrm{cor}} \sqrt{ \left[ \frac{\sigma_{a}}{ \sum\limits_j I_{j}} \right]^2+ \left[\frac{\sigma(T_r)}{T_r}\right]^2 + \left[\frac{\sigma(f_s)}{f_s}\right]^2 }
		\end{equation}
		
		\subsubsection{Time and thickness corrections: TI and TH }\label{sc:TI}\label{sc:TH}
		
			The time and thickness corrections are nearly identical to the transmission and flux corrections (\S\ref{sc:TR}) and equally straightforward: the detected intensity is proportional to the measurement time\footnote{With the notable exception of image plates which suffer from competitive decay. This decay  should be corrected for before this step as discussed in \S \ref{sc:GA}}, and the amount of scattered radiation is proportional to the amount of material in the beam. The thickness correction is applied to correct for differences in the amount material in the beam. 

		These two corrections are applied through normalisation of the measured intensity with the sample measurement time $t_s$ and thickness $d_s$:
		\begin{equation}\label{eq:TH:I}
			I_{j,\mathrm{cor}}= \frac{I_j}{t_s d_s} 
		\end{equation}

		\begin{equation}\label{eq:TH:sa}
			\sigma_{a,\mathrm{cor}}=  \sum\limits_j I_{j,\mathrm{cor}} \sqrt{ \left[ \frac{\sigma_{a}}{ \sum\limits_j I_{j}} \right]^2+\left[\frac{\sigma(t_s)}{t_s}\right]^2 + \left[\frac{\sigma(d_s)}{d_s}\right]^2 }
		\end{equation}

		\subsubsection{Absolute intensity correction: AU }\label{sc:AU}
		
			Scaling the data to reflect the materials' differential scattering cross-section can be a great boon to the value of the data. This scaling allows for the evaluation of the scattering power of the sample in material specific absolute terms which can lead to e.g. the determination of the volume fraction of scatterers or their specific surface area \emph{and} to check the validity of assumptions made. Furthermore, it allows for proper scaling between techniques, and can help distinguish multiple scattering effects. This scaling gives the scattering profile the units of scattering probability per unit time, per sample volume, per incident flux and per solid angle, which if worked out comes to m$^{-1}$sr$^{-1}$ (though centimetres are sometimes used instead of metre) \cite{Wignall-1987,Boesecke-1997,Dreiss-2006,Zhang-2010a}. 
		
		This scaling can be achieved in two ways; either through direct calibration with samples whose scattering power can be calculated, or through the use of secondary standards. A discussion of the benefits and drawbacks of either have been well explained by \citet{Dreiss-2006}. The most straightforward method is using a secondary standard such as Lupolen or calibrated glassy carbon samples \cite{Russell-1988,Zhang-2010a}, as they do not require detailed knowledge on detector behaviour and beam profiles, and scatter significantly allowing for rapid collection of sufficient intensity to perform the calibration. The scattering of these samples in absolute intensity units is determined beforehand, and its calibrated datafile should come with the sample \cite{Wignall-1987,Zhang-2010a}. By comparing the intensity in the calibrated datafile with the locally collected intensity, a calibration factor $C$ can be determined:
		
		\begin{equation}
			C=\frac{\left(\frac{\delta \Sigma}{\delta \Omega}\right)_{st}}{I_{st,\mathrm{cor}}}
		\end{equation}
		where the subscript $_{st}$ denotes the calibration standard, and $I_{st,\mathrm{cor}}$ is the measured and corrected intensity and $\left(\frac{\delta \Sigma}{\delta \Omega}\right)_{st}$ is the known scattering pattern (calibrated datafile) from the calibration sample. The calibration factor can be determined by a least-squares fit or linear regression (with least squares allowing for inclusion of counting statistics and optional flat background contribution). 
		
		The accuracy of this determination depends on a large amount of uncertainties. Practically, though, an accuracy of about 10\% can be achieved \cite{Mylonas-2007}. It may be approximated as:
		\begin{equation}\label{eq:AU:sC}
			\left[\frac{\sigma(C)}{C}\right]^2= \left[ \frac{\sum\limits_{j} \sigma\left(\left(\frac{\delta \Sigma}{\delta \Omega}\right)_{st,j}\right)}{\sum\limits_{j} \left(\frac{\delta \Sigma}{\delta \Omega}\right)_{st,j}} \right]^2+\left[\frac{\sum\limits_j \sigma_a}{\sum\limits_j I_{j}}\right]^2
		\end{equation}
		
		The approximation is finally applied to the measured data as:
		\begin{equation}\label{eq:AU:I}
			I_{j,\mathrm{cor}}= I_j C
		\end{equation}
		Whose absolute uncertainty follows:
		\begin{equation}\label{eq:AU:sa}
			\sigma_{a,\mathrm{cor}}= \sum\limits_j I_{j,\mathrm{cor}} \sqrt{ \left[ \frac{\sigma_{a}}{\sum\limits_j I_{j}} \right]^2+ \left[\frac{\sigma(C)}{C}\right]^2 }
		\end{equation}
		
		\subsubsection{Background correction: BG }\label{sc:BG}
		
			In any scattering measurement, it is of great importance to isolate the (coherent) scattering of the objects under investigation from all other parasitic scattering contributions such as windows, solvents, gases, collimators, sample holders and any incoherent scattering components. For example, this would be the removal of the scattering of capillaries and solvents from the scattering pattern of a suspension or solution, or the removal of instrumental background (scattering from windows, air spaces, etc.) from the scattering pattern collected for a polymer film. A background measurement thus contains as many as possible of the components present in the sample measurement, minus the actual sample. A detailed discussion of suitable background samples can be found in  \citet{Brulet-2007}. 
			
		In most cases, the background measurement should be performed with as many variables identical to the sample measurement, and subjected to the same data corrections\footnote{The background thickness correction for measurements where the background measurement consists of no sample at all (f.ex. backgrounds for film samples or sheets) is slightly special. The ``thickness'' of this background measurement should be set identical to the thickness of the sample in the sample measurement.}. This means that the background measurement should be measured for the same amount of time as the sample measurement. However, in case the detector behaviour is well characterised and the signal-to-noise ratio (c.q. sample-to-background scattering signal) is large, the sample-to-background measurement time ratio may be skewed (to favour sample measurement time) in order to improve the statistics after background subtraction \cite{Steinhart-1993,Pauw-B2012}. The correction is applied as:
		
		\begin{equation}\label{eq:BG:I}
			I_{j,\mathrm{cor}}=I_j - I_{j,b} 
		\end{equation}
		 with $I_{j,b}$ the background measurement intensity for datapoint $j$. The uncertainties, both absolute and relative, propagate as follows:
		\begin{equation}\label{eq:BG:sr}
			\sigma_{r,j,\mathrm{cor}}=\sqrt{ \sigma_{r,j}^2+\sigma_{r,j,b}^2}
		\end{equation}
		\begin{equation}\label{eq:BG:sa}
			\sigma_{a,\mathrm{cor}}=\sqrt{ \sigma_{a}^2+\sigma_{a,b}^2}
		\end{equation}
				
		\subsubsection{Correcting for spherical angles: SP }\label{sc:SP}
		
\begin{figure}
		   \centering
		   \includegraphics[angle=0, width=0.60\textwidth]{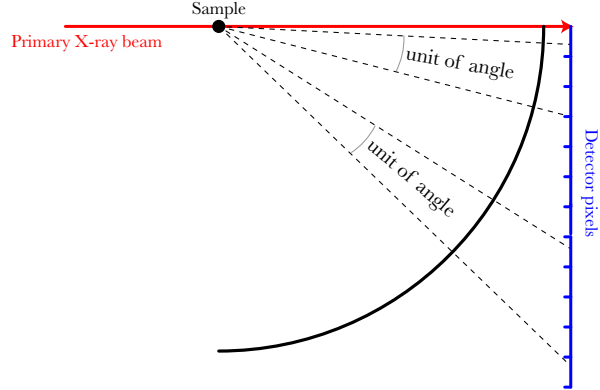} 
		   \caption{The need for spherical corrections illustrated for straight detectors (as opposed to tilted detectors). One unit angle covers a different number of pixels, which needs to be corrected for. \label{fg:SP}}
		\end{figure}

			Most detectors are flat with uniform, square pixels, but we wish to collect the intensity over a solid angle of a (virtual) sphere. The projection of the detector pixels on the sphere results in a difference in solid angle covered by each pixel (illustrated in Figure \ref{fg:SP}) \cite{Boesecke-1997,Barna-1999,Le-Flanchec-1996}. Therefore, we need to correct the intensity for the difference between these areas\footnote{This is further exacerbated if the detector is tilted with respect to the beam, and thus has a ``point of normal incidence'' with respect to the sample which differs from the direct beam position.}. 
		
		The correction for this effect achieved by means of a few geometrical parameters. This correction is given by \cite{Boesecke-1997} as:
		\begin{equation}\label{eq:SP} 
			\frac{L_P^2}{p_x p_y}\frac{L_P}{L_0}
		\end{equation}
		where $L_P$ is the distance from the sample to the pixel, $L_0$ the distance from the sample to the point of normal incidence (usually identical to the direct beam position except in case of tilted detectors), and $p_x$ and $p_y$ are the sizes of the pixels in the horizontal and vertical direction, respectively. As it is unnormalised, this correction factor typically assumes very large values. When normalised to assume a value of 1 at the point of normal incidence, the correction becomes:
		\begin{equation}\label{eq:SP:I}
			I_{j,\mathrm{cor}}=I_j \frac{L_P^3}{L_0^3}
		\end{equation}
		
\begin{figure}
			\centering
   			\includegraphics[angle=0, width=0.50\textwidth]{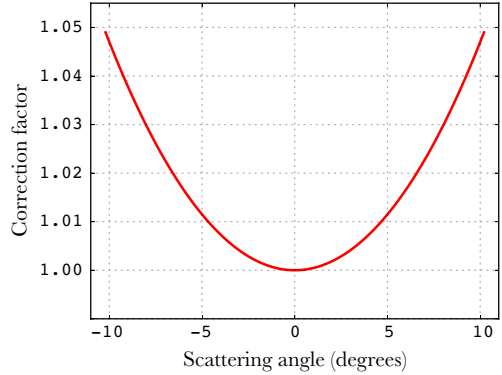} 
   			\caption{Area dilation correction showing an increasing need for application of the correction beyond about 5 degrees.}\label{fg:SPeff}
		\end{figure}

		Its magnitude is shown in Figure \ref{fg:SPeff}, and is generally less than 1\% for scattering angles lower than 5 degrees. It very quickly becomes more severe beyond those angles. 
		%WORK HERE
		
		\subsubsection{Sample self-absorption correction: SA }\label{sc:SA}
			
			When scattering occurs in a sample, the scattered radiation has to travel some distance through the sample. Depending on the sample geometry and the scattering angle, this scattered radiation has to travel through more or less material. The direction-dependent absorption thus occurring can induce an angle-dependent scattered intensity reduction which is most severe for scattering to wider angles and for samples with a high attenuation coefficient \cite{Brulet-2007,Sulyanov-2012,Zeidler-2012,Bendert-2013}. This is essentially a correction of the transmission factor correction described in \S \ref{sc:TR}.

		Its correction for plate-like samples to a scattering pattern takes the form of:
		\begin{equation}\label{eq:SA:I}
%			I_{j,\mathrm{cor}}=I_j \frac{ 1-T^{\left[\frac{1}{\cos(2\theta)}-1\right]} }{ \ln( T )-\frac{1}{ \cos( 2 \theta ) } \ln ( T )}
			I_{j,\mathrm{cor}}=\left\{
			\begin{array}{lr}
				I_j \frac{ 1-T^{\left[\frac{1}{\cos(2\theta)}-1\right]} }{ \ln( T )-\frac{1}{ \cos( 2 \theta ) } \ln ( T )} & \textrm{, for } 2\theta\neq 0 \\
				I_j & \textrm{, for } 2\theta=0 \\	
			\end{array}
			\right.
		\end{equation}
		which can be expressed in terms of linear absorption coefficient $\mu$ and thickness $d$ as:
		\begin{equation}\label{eq:final}
			I_{j,\mathrm{cor}}=\left\{
			\begin{array}{lr}
				I_j \exp(\mu d) \frac{-\exp(-\mu d) +\exp\left(-\mu d/\cos(2\theta)\right)}{\mu d - \mu d/\cos(2\theta)} & \textrm{, for } 2\theta\neq 0 \\
				I_j & \textrm{, for } 2\theta=0 \\				
			\end{array}
			\right.
		\end{equation}
		where $2\theta$ denotes the scattering angle. As the numerator and denominator of the fraction tend to zero for $2\theta=0$, at that point $I_{j,\mathrm{cor}}=I_j$ must be substituted. 
		This correction is only valid for plate-like samples, for which it is still straightforward to derive. For spherical samples and cylindrical samples, the direction-dependent attenuation becomes much more complicated \cite{Sulyanov-2012,Zeidler-2012}, and an extra level of difficulty is added for off-center beams \cite{Bendert-2013}. %also about cylindrical samples is Paalman-1962, Dwiggins-1971, 
		
\begin{figure}
			\centering
   			\includegraphics[angle=0, width=0.95\textwidth]{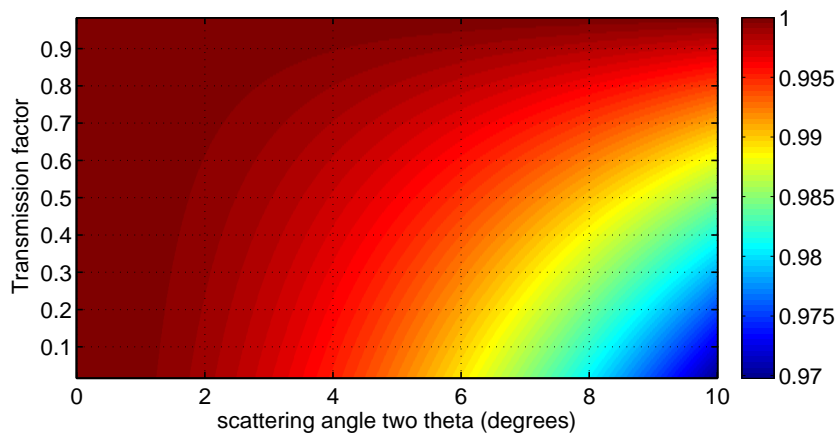} 
   			\caption{Absorption due to sample geometry for a sample for a range of absorptions}\label{fg:SA}
		\end{figure}

		Figure \ref{fg:SA} shows the magnitude of the correction depending on the transmission factor and scattering angle. As previously remarked, the effect is most severe for highly absorbing samples and wide angles. 

		As the correction is rather minimal for small-angle scattering, its effects on the uncertainties are expected equally negligible. Given the estimated complexity of the uncertainty propagation in this case, its derivation is here omitted.

		\subsubsection{Multiple scattering correction: MS }\label{sc:MS}
		
			Multiple scattering occurs when a scattered photon still travelling through the material undergoes a subsequent scattering event. As the probability for any photon to scatter (irrespective of whether it has scattered or not) is proportional to the scattering cross-section of the material and the amount of sample in the beam, multiple scattering becomes more dominant for strongly scattering, thick samples \cite{Schelten-1980,Copley-1988,Mazumder-1992,Mazumder-2003}. It effects a ``smearing'' of the true scattering profile, which can significantly affect analyses \cite{Chonacky-1969,Copley-1988}. 
			
			When the possibility of multiple scattering exists for a particular sample measurement (i.e. with a transmission factor below approximately $1/e$ and strongly scattering samples) it is prudent to test whether it is a significant contribution. This can be performed experimentally by measuring samples with different thicknesses or by changing the incident wavelength. If the scattering profile after corrections significantly differ, chances are that multiple scattering may need to be accounted for \cite{Mazumder-2003}. Alternatively, the multiple scattering effect can be estimated analytically \cite{Schelten-1980} or using Monte-Carlo based procedures \cite{Copley-1988,Seeger-2003}
			
			Like any smearing effect, correcting (also known as ``desmearing'') data for multiple scattering effects is much more involved than smearing the model fitting function. When given the choice, implementing a smearing procedure in the fitting model rather than the data is preferred \cite{Ghosh-2012}. Correcting for multiple scattering is generally a complex, iterative procedure where the multiple scattering smearing profile is estimated and removed from the data \cite{Mazumder-2003}. It becomes even more complicated for samples with direction-dependent sample thicknesses and hence different multiple scattering probabilities \cite{Blech-1965,Strunz-2000}. One avenue for simplifying the correction and estimation is by approximation of the multiple scattering effect as mainly consisting of double scattering \cite{Chonacky-1969,Ghosh-2012,Bendert-2013}.
			
		\subsubsection{Instrumental smearing effects correction: SM }\label{sc:SM}
			
			The incident beam characteristics (in particular its profile and wavelength spread) and detector position sensing inaccuracies cause a smearing of the detected scattering pattern \cite{Pedersen-1990,Barker-1995,Harris-1995}. Apart from the wavelength spread, the smearing contributions can be evaluated as the image of the direct beam on the detector with which the ``true'' scattering convolves \cite{Pedersen-1991}. The wavelength-smearing effect of crystal reflection-monochromatised radiation is typically considered small in comparison to the other smearing contributors.
			
			Such corrections are usually not applied for pinhole-collimated X-ray scattering instruments, where if they are considered at all they are usually incorporated as a model smearing rather than a data desmearing. There are some notable exceptions by \citet{Le-Flanchec-1996} and \citet{Stribeck-2008}, in the latter example it is applied to allow for improved intercomparability of 2D scattering patterns collected with differing collimation. 
			
			Beam desmearing corrections are more commonly applied for instruments with line-collimated beams (e.g. instruments discussed in \S \ref{sc:slit} and \S \ref{sc:bh}), though model smearing rather than data desmearing is still recommended \cite{Gravatt-1969,Ruland-1978}. As slit-smeared instruments have existed as long as small-angle scattering itself, desmearing procedures are available of many types and vintages. Some notable ones include \citet{Lake-1967,Strobl-1970,Vonk-1971,Glatter-1974} and \citet{Singh-1993}. One commonly implemented iterative desmearing procedure is described by \citet{Lake-1967}\cite{Vad-2011}. The disadvantages of any desmearing procedure are their tendency to amplify small differences leading to increased noise levels, and their arbitrary cut-off criterion \cite{Strobl-1970}. The former disadvantage is partially offset by the improved data accuracy of the initial data (due to the increased flux of slit-collimated instruments), and the second can be overcome through introduction of cut-off criteria \cite{Vad-2011}.
			
		\subsubsection{Data binning }\label{sc:int}
			
			At some point in the data correction procedure for isotropically scattering samples, a data reduction step is performed, known as ``integration'', ''averaging'' or ``binning''. For isotropically scattering samples, a reduction in dimensionality of the data usually accompanies this procedure (e.g. from 2D images to 1D plots), by grouping and averaging pixels with similar scattering angle $q$ irrespective of their azimuthal angle on the detector (denoted $\psi$). For anisotropically scattering samples pixels with similar $q$ \emph{and} $\psi$ can be combined to form a new 2D dataset but with a reduced amount of datapoints \cite{Pauw-2010a,Pauw-A2013}, though some dispense with binning altogether \cite{Paul-2013}. 
			
			The advantages of this step are threefold. Firstly, the data becomes more manageable, allowing for example faster fitting and improved data visualisation. Secondly, the relative data uncertainties become smaller for the averaged data. Lastly, the standard deviation between similar pixels in a group (c.q. bin) can provide a good estimate for the actual uncertainty on the average value if this standard deviation exceeds the photon counting statistics-based estimate propagated until this step\footnote{The photon counting (Poisson) statistics defines the absolute minimum possible uncertainty in any counting procedure. It does not consider other contributors to noise such as the variance between pixel sensitivities or electronic noise.}.
			
			More specifically \cite{Pauw-2013}: for radial averaging the many datapoints $I_j$ collected from each pixel on the the detector are reduced into a small number of $q$-bins $I_{\mathrm{qbin}}$ before the data analysis procedures. In this reduction step, each measured datapoint collected between the bin edges (class limits) $q_n$ and $q_{n+1}$ is averaged and assumed valid for the mean $\overline{q}=\langle q\in [q_n, q_{n+1}] \rangle$, i.e.: 
			\begin{equation}\label{eq:Iint}
				I_{\mathrm{qbin}}(\overline{q})=\langle I_j (q\in [q_n, q_{n+1}]) \rangle
			\end{equation}
			
			\begin{equation}\label{eq:int:sr}
				\sigma_{r,qbin,\mathrm{cor}}=\max \left\{
				\begin{array}{l}
					\frac{1}{N_\mathrm{qbin}}\sqrt{ \sum\limits_{q\in [q_n, q_{n+1}]} \sigma_{r,j}^2} \\
					\sqrt{\frac{1}{N_\mathrm{qbin}-1}\sum\limits_{q\in [q_n, q_{n+1}]}\left(I_j-I_{\mathrm{qbin}}\right)^2} \rule{0pt}{7.5mm }\\
				\end{array}
				\right.
			\end{equation}
			where the summation is over all datapoints $j$ falling within the bin edges, $N_\mathrm{qbin}$ the total number of datapoints in the bin. As previously mentioned and evident from equation \ref{eq:int:sr}, the maximum value is chosen between the propagated uncertainty and the sample standard deviation in the bin. In other words, if the sample standard deviation of the pixels in the bin exceeds the estimate based on the previously propagated uncertainty, the sample standard deviation is considered a more accurate estimate. This can be further augmented to never have a relative uncertainty estimate less than 1\% of the intensity, as it is (even with the most stringent corrections) challenging to get more accurate than this \cite{Hura-2000}.

			There is still a choice to be made in this procedure, that is the spacing between the bin edges. Normally, this is chosen either uniform or logarithmically spaced (with more data points at low values) \cite{Ilavsky-2012}. However, for data with sharp features, a more involved choice might be preferred \cite{Sanders-2001,Cervellino-2006}.

	\subsection{The order of corrections for a standard sample}
\begin{figure}
			\centering
			\includegraphics[angle=0, width=0.4\textwidth]{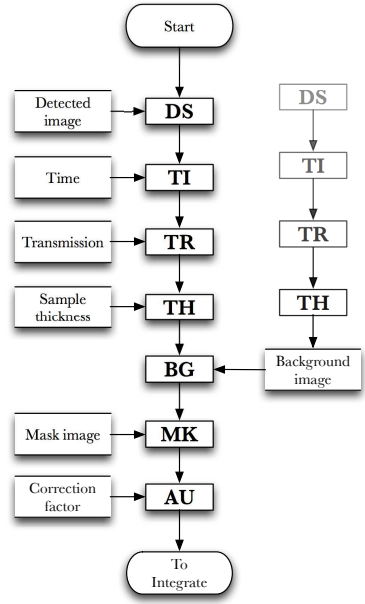} 
			\caption{Minimum amount of corrections suitable to samples without a strongly absorbing container, with a stable x-ray beam, a good detector and low dark current c.q. natural background. \label{fg:minimalseq}}
		\end{figure}

		The absolute minimal number of corrections (Figure \ref{fg:minimalseq}) to apply consist of the normalisations to time, transmission and thickness and subtraction of the background. This works reasonably well for strongly scattering samples with low absorptions, without strong absorbance from the container. Furthermore it requires a problem-free detector and a stable X-ray source. 
		
\begin{figure}
			\centering
			\includegraphics[angle=0, width=0.85\textwidth]{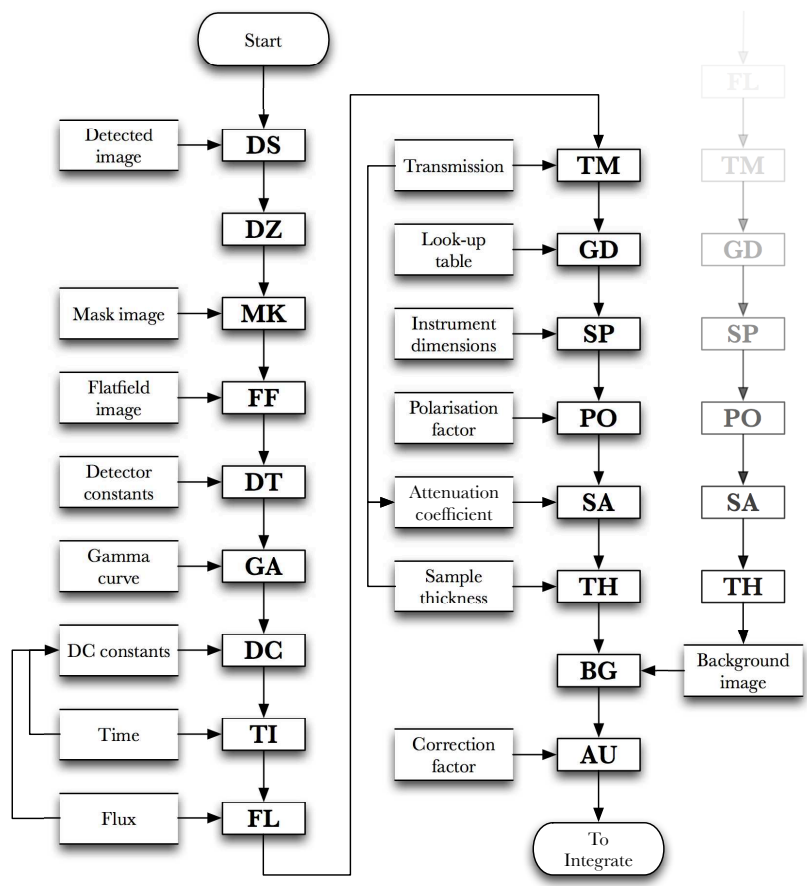} 
			\caption{Standard sequence of corrections suitable to samples without a strongly absorbing container. \label{fg:basicseq}}
		\end{figure}

		The standard set of corrections are a little more involved but allow for more flexible experimental conditions (Figure \ref{fg:basicseq}). Strongly absorbing samples, samples with low scattering power and instruments with imperfect detectors (CCD's, image plates and wire chambers) are supported by this scheme. Samples contained in a strongly scattering and/or absorbing container, however, are not supported by this scheme. Its application to such samples would lead to an incorrect estimation of the absolute scattering power from such samples. In case the sample container shows appreciable scattering, this scheme furthermore leads to incorrect background subtraction.

		The corrections described above work reasonably well for most samples. There is, however, one more level of difficulty in the search for perfection when working with samples in containers (e.g. capillaries, or other container-sample-container sandwiches). The challenge with these is that the incident radiation first encounters an amount of absorbing and scattering container material, \emph{then} passes through the absorbing and scattering sample, upon which it \emph{again} passes through an amount of absorbing and scattering container material\footnote{With yet another challenge created by capillaries, as their diameter and wall thickness is not all that well defined, and an off-centered beam would make direction-dependent absorption corrections unwieldy.}. 
		
		In advanced corrections, suitable for most samples imaginable, one would need to apply to the scattering image:
		\begin{enumerate}
			\item{}A background correction for the scattering from the upstream container wall, corrected for direction dependent absorption by the upstream container wall, the sample as well as the downstream container wall.
			\item{}Corrections for the actual sample thickness, incident flux reduced by the absorption of the first container wall and direction-dependent absorption by both the sample as well as the downstream container wall.
			\item{}A background correction for the scattering (now from a primary beam reduced in intensity by the absorption from the upstream container wall and the sample) from the downstream container wall and its direction-dependent absorption.
		\end{enumerate}
		Such a scheme would require the splitting of scattering intensity of the sample container and a non-self-scattering direction-dependent absorption correction. Work in that direction has been shown by (amongst others) \citet{Brulet-2007}.
		
	\subsection{The development of reduced data storage standards}\label{sc:universalstorage}
		Besides efforts to store the raw collected data in archival formats currently underway at some of the larger institutions, there has also been some development in storing the data obtained after application of all these corrections in a universal (archival) format. These can be separated into two categories: the storage of integrated data (1D), and the storage of data of higher dimensionality (2D or more). Both formats should allow for the storage of accompanying metadata.
		
		The opinions on the storage type of corrected (and integrated) 1D small-angle scattering data is roughly divided into two factions. The most common format for exchange and storage of such data is as a human-readable file (in either ASCII or UTF-8 encoding) consisting of a header containing the metadata, and the body containing the corrected data commonly in scattering vector Q, scattering cross-section I and the estimated uncertainty on the latter \cite{Jacques-2012}. The benefits of this storage method is that it is easily understood and accepted by users and programs alike. It is furthermore one of the easiest formats to write for the scientist-cum-programmer. The disadvantage is that it is an ill-defined, ad-hoc standard, which may or may not contain all essential information in the header. While the sasCIF effort set out to alleviate some of these issues, its current state is unknown \cite{Malfois-2000}.
		
		The second corrected 1D data storage type has recently emerged from a lengthy development process in collaboration with the small-angle scattering community. This ``canSAS 1D'' format is an XML-based data storage format, acting as a flexible but well-defined container that can accommodate a large variety of data \cite{cansas-2013}. The disadvantage is the necessity to write in an XML-based format which is not overly complicated but requires a modicum of effort to implement. The adoption of this standard is slow but gradual.
		
		The storage of corrected multidimensional (2D and higher) SAS data is surprisingly less bifurcated. The general consensus in the community is that simple image file types are insufficient to encapsulate all the details of the corrected data, and that a hierarchical data format such as that provided by the HDF5 format is required \cite{Folk-2010}. Building upon the base HDF5 structure is the NeXus format for storing (raw) instrument data at large facilities \cite{Klosowski-1997,Konnecke-2006}, and building upon \emph{that} (or at least offering compatibility) is the SAS-specific ``canSAS2012'' format \cite{cansas-2013}. This allows for the storage of any data, of which a subset is required data (in particular the scattering intensity and orthogonal scattering vectors). As there are but very few programs available capable of analysing 2D data, its adoption rate cannot be estimated.

\section{What's next? A few words on data fitting}
After correction of the collected data to obtain the scattering from the substance of interest (and only that substance), the data can be subjected to data analysis to extract relevant physical parameters. Such data analyses are heavily dependent on the material and structure in question, and very few general paths to answers exist. However, similar structures and materials often require similar data analysis pathways, so that these may be used as guidance to develop the analysis methodology for a novel material. 

In the early days --- without the convenience of near-obscene amounts of computing power --- data analysis mainly revolved around severe assumptions allowing for data linearisation. Examples of these are still found, with their linearised data often denoted as ``Guinier plots'' \cite{Guinier-1955} (also developed by several others \cite{Fankuchen-1944}), ``Debye-B\"uche plots'' \cite{Debye-1949}, ``Kratky plots'' \cite{Glatter-1982} and ``Porod plots'' \cite{Glatter-1982}, to name but a few. While these linearisations may have some value for rough evaluation of data, they should never be relied on as the final analysis. One of the major drawbacks of linearisation of data is the visual skewing of the datapoint weights. Especially if no data uncertainty information is available, the linearisation will put a heavy emphasis on either the initial (``Guinier'') or the latter datapoints (``Porod''), and forcibly effect a linear interrelationship. They furthermore often rely on data which either has a high probability of distortion (such as the Guinier plot, which can be easily affected by structure factor effects, smearing or parasitic scattering) or a very low signal-to-noise ratio (such as the Porod plot, which is highly reliant on accurate background subtraction and minimisation of the instrumental scattering). If and when these plots are exploited to obtain numbers, ensure that the related fits are performed on the original data, not on the linearised data. The use of modernised variants is recommended, which are available for some of the aforementioned relationships. These modernised variants may be capable of fitting an extended region of data \cite{Beaucage-1996,Hammouda-2010}. 

A much more appropriate data fitting methodology came about following the spread of computers. Least-squares fitting of data to appropriate models can deliver better quality results from data. If the data is complete with accurate uncertainty estimates, models can be evaluated on their descriptive ability and over fitting can be prevented. A good treatise on this methodology is given by \citet{Pedersen-1997}. These fitting methods typically consist of a combination of functions describing each of the three previously described morphological aspects: 1) a ``Form Factor'' function describing the elementary scatterer shape (the intraparticle scattering contribution), 2) a size distribution function descriptive of the dispersity in size of the elementary scatterers, and 3) a ``Structure Factor'' which describes the interparticle scattering contribution. While more flexible than their predecessors (the linearised approximations), these  models do require assumptions to be made on all three morphological aspects (c.f. \S \ref{sc:ssa}). For example, work by \citet{Abecassis-2007} assumes diluteness (i.e. a uniform ``Structure Factor'' contribution equalling 1), a Gaussian-shaped size distribution of scatterers, and a spherical shape for the scatterers.

As the assumption of \emph{two} of the morphological aspects fixes the third to a single unique solution, the most modern data analysis methods revolve around the form-free retrieval of the unassumed aspect. When possible, these methods should be preferred as they reduce the number of assumptions made, and may be more straightforward to fully describe the data with, and may be easier to justify the assumptions of (as only two, not three assumptions have to be justified). As several of these methods have been mentioned in \S \ref{sc:ssa}, they are not repeated here.

Whichever route is chosen for analysis, there are a few data analysis pitfalls to be wary of:
\begin{itemize}
\item{\textsc{Data linearisation}: } As noted by others, data linearisation has become unnecessary in current-day analyses \cite{Stribeck-2002}. They can be used as a quick evaluation of the data, but their use as a final recourse should be discouraged. In particular the actual graphical analysis of linearised data is a practice that should be retired. The most extreme cases, where the linearised data is not linear yet still subjected to graphical analyses, are occasionally encountered \cite{Das-2005,Wang-2008}. When linearisations are used to extract parameters, they should at least exhibit a linear region (as reiterated by \citet{Porod-1951}).
\item{\textsc{Uniqueness ``it fits, so it must be true''}: } As indicated in \S \ref{sc:ssa}, there are a large number of solutions that would fit a given scattering pattern, necessitating assumptions for two of the three morphological aspects of packing, polydispersity and shape. The assumptions should ideally be supported with supplementary techniques such as microscopy or porosimetry, or from fundamental considerations of the emergence of the scatterers (as expressly stated by \citet{Evrard-2005}).

In standard least-squares fitting procedures, however, assumptions have to be made on all three aspects which often leads to imperfect descriptions of the scattering data. It may then transpire that, once a combination of aspects has been found that fits the data, it is thought that these must be the correct aspects as they describe the data (a form of circular reasoning similar to the logical fallacy of ``begging the question''). In other words, quoting Dr. J. Ilavsky: ``The fact that A model fits your data is NOT proof that it is THE appropriate model''. While this fallacy is not always clearly indicated, some papers do indicate that the success of a particular structural model to describe the data is evidence for its validity \cite{Tekobo-2011,Parent-2011,Meli-2012}. As mentioned, the choice of any particular model should be supported by supplementary information.

\item{\textsc{Lack of uncertainties and overly optimistic fits}: } 
Uncertainties allow for the weighting of the data to its uncertainty, so that accurate datapoints weigh more heavily in least-squares optimisation than datapoints with large uncertainties. Furthermore, the provision of uncertainties allows for determination of a goodness of fit value which indicates whether or not the model fits the data (on average) to within the uncertainty of the data \cite{Pedersen-1997,Pauw-2013}. A lack of uncertainties does not allow any further evaluation of a model than an estimation by eye, whose analysis capabilities are easily swayed by the choice of axes and datapoint size. Coupled with a modicum of ``wishful thinking'', this may lead to overly optimistic fits that only coarsely describe the data. Such fits still provide structural parameters, but their veracity is dubious. An example of optimistic data fitting is given by \cite{White-2010}, with poor fits to data with unknown uncertainties. Fortunately, their results appear to agree with TEM data. 

\item{\textsc{Unsuitable range}: } A scattering pattern is metrologically limited to a finite angular range with equally limited angular steps. It can be shown that the smallest measurable feature is closely related to the largest measurable angle through $R_\mathrm{min}\approx\pi/\max(q)$ \cite{Pauw-2013}. The largest measurable feature is ultimately limited to the angular divergence of the incident beam (as can be easily derived from considerations of the Bonse-Hart-type diffractometer). It can be approximated as $R_\mathrm{max}\approx\pi/w$, where $w$ is the width of the beam profile on the detector in $q$. It will in most cases be further limited by the smallest step size in $q$ of the detection system. When the determined structural features from data analysis contain information on sizes beyond these limits, a check for their evidence in the measured data must be made. 

\end{itemize}

Through recognition of these pitfalls as well as a thorough understanding of the limitations and applicability of the applied models, reliable data analysis may be achieved. To aid the process of fitting of data, many software packages have become available over time, a few of which should be considered. Commonly used packages (for non-biological systems) are Irena \cite{Ilavsky-2009}, SASfit \cite{SASfit} and Scatter \cite{Foerster-2010}. There is furthermore a large set of tools available in the ATSAS package \cite{Petoukhov-2012}. To ease the troubles of software installation, web-based analysis tools have recently become available such as for the previous ATSAS package and a Bayesian inverse Fourier transform routine by \citet{Hansen-2012}. 

\section{Conclusions}
The improvements in the small-angle scattering instrumentation have recently enabled easy collection of data from a large variety of ex-situ and in-situ studies. This data theoretically contains a wealth of information on the nanostructure in the sample, the scope of which is best illustrated by the (rapidly increasing) number of publications applying small-angle scattering to a great number of fields. Its elucidative power thus exemplified, it is time to ensure that the data that forms the basis of these interpretations is of the highest quality, so that the conclusions are sound and authoritative.

It is our hope that the comprehensive set of data corrections provided herein (with consistent equations for the correction as well as the uncertainty propagation) can be a step towards this goal. While most corrections have details that have necessarily been left out, the information given may provide the insight required to determine which of the corrections are required, at what stage, to what accuracy and at what cost of programmatical complexity. 

The casual small-angle scattering user should expect to get accurate data (subjected to the most stringent corrections) from the instrument responsible, and should never have to implement corrections. Ideally, the user would also be able to confer with a data analysis expert on the best analysis methodology to apply. If and when this comes to pass, it should never be forgotten that underneath all the gloss lies an instrument made from common nuts and bolts, that data is trimmed and adjusted to remove --- as much as possible --- the nuts and bolts from the equation, and that in the nuts and bolts lie the limitations of the technique. Understanding of the nuts and bolts, the corrections, the analyses and their limitations, is key to understanding the final results that pop out when the machine goes ``Ping!''.

\section{Acknowledgements}
With the current deluge of publications on even the remotest of topics, and the general lack of search and indexing facilities to provide a suitable response, it is likely that a good paper has not been mentioned. While the author has taken special care to prevent such omissions (and apologises in advance for omitting), suggestions for papers for inclusion in future work are always welcomed. 

This work has come to fruition through discussions with a large number of remarkable people. The author therefore thanks (in alphabetical order): Ingo Bre\ss{}ler, Ron Ghosh, Martin Hollamby, Jan Ilavsky, Pete Jemian, J\'er\^ome Kieffer, Kell Mortensen, Yojiro Oba, Masato Ohnuma, Jan-Skov Pedersen, Adrian Rennie, Julian Rosalie, Kenji Sakurai, Hossein Sepehri-Amin, Masaaki Sugiyama, Samuel Tardif, Andreas Th\"unemann and coworkers, and Thomas Zemb and coworkers. 

%Ingo Bressler, Ron Ghosh, Martin Hollamby, Jan Ilavsky, Pete Jemian, J\'er\^ome Kieffer, Kell Mortensen, Yojiro Oba, Masato Ohnuma, Jan-Skov Pedersen, Adrian Rennie, Julian Rosalie, Kenji Sakurai, Hossein Sepehri-Amin, Masaaki Sugiyama, Samuel Tardif, Andreas Th\"unemann and coworkers, and Thomas Zemb and coworkers. 

\section{Appendix: Measurement checklist}
For the reduction and correction algorithms, one has to ensure the following is known:

The geometric information:
\begin{itemize}
\item{$\square$ } The sample-to-detector distance
\item{$\square$ } The wavelength and its expected degree of monochromaticity. If the photon energy $E$ is supplied in units of $keV$, this can be converted to \AA ngstr\"om (1 \AA $=10^{-10}$ m) through $\lambda(\AA)=12.398 / E(keV)$ (conversion factor from the 2002 NIST CODATA database)
\item{$\square$ } The position of the direct beam on the detector (in pixels)
\item{$\square$ } The point-of-normal-incidence in pixels for tilted detectors (i.e. not mounted perpendicular to the direct beam)
\item{$\square$ } The collimator sizes, types and distances between the elements (for publication and check for maximum transversal coherence length).

\end{itemize}

The detector information:
\begin{itemize}
\item{$\square$ } The detector name
\item{$\square$ } The number of pixels in the horizontal and vertical directions
\item{$\square$ } The size of the individual detector pixels in horizontal and vertical directions (in meters)
\item{$\square$ } The detector file data type (f.ex. 16-bit unsigned integers)
\item{$\square$ } The detector file endianness
\item{$\square$ } The required image transformation to transform the detector output image to the laboratory frame of reference
\item{$\square$ } The detector rotation offset in case of a detector rotated with an arbitrary number of degrees
\end{itemize}

The correction information:
\begin{itemize}
\item{$\square$ } The filename of the mask image with the masked pixels
\item{$\square$ } The mask acceptance window for valid pixels (analog to a bandpass filter with a low intensity cut-off and a high-intensity cut-off)
\item{$\square$ } The flatfield image filename (if applicable)
\item{$\square$ } The darkcurrent components (time-independent, time dependent and flux-dependent)
\item{$\square$ } Details on the gamma correction, or information on the region of linearity (and the maximum deviation from this linearity)
\item{$\square$ } The detector geometric distortion look-up table (if applicable)
\item{$\square$ } The absolute intensity standard sample identifier
\item{$\square$ } The absolute intensity calibration factor
\item{$\square$ } Details on the polarisation of the beam
\end{itemize}

Note the ion chamber and pin-diode amplifier settings (and readings for a ``normal'' measurement without a sample in the beam) for troubleshooting ease and transmission calculation. Also ask for motor movement limits and positioning accuracy. Lastly, write down anything that appears to be valuable information during the measurement session (this naturally includes \emph{everything} the instrument responsible tells you).

Finally, check if all of the beamline computers are set to the same time and date, and that these are correct.

\subsection{The user should measure the following}
\subsubsection{Transmission measurement}
For each measurement (indeed, each datafile), the average transmission factor for the duration of the measurement has to be calculated (e.g. for correct background subtraction). The methods for this have been given before in \S \ref{sc:TR}. The on-line measurement techniques allow for constant collection of the incident beam flux as well as the transmission factor during the measurement (often with a frequency of several Hertz), which should also be stored. Deviations in the transmission factor during the measurement are a good indication of sample instability or motion. 

\subsubsection{Background measurement}
The background corrections have been discussed in detail in \S \ref{sc:BG}. Repeat the background measurement for each change in geometrical configuration or change of solvent or capillary type and size. 

\subsubsection{Sample measurement}
The actual sample measurement should be measured long enough for collection of sufficiently accurate intensities, and should be measured multiple times in sequence to check for sample instability (and possible de-zingering). Multiple samples of the same material should be measured to determine the statistical uncertainty of the physical parameters resulting from the eventual pattern analysis. For dynamic systems this is not always possible, but repeated runs of the same dynamic system should provide some insights on the final uncertainties. 

\subsubsection{Darkcurrent measurement}
If the details of the dark current components are not known (c.f. \S \ref{sc:DC}), measure a new darkcurrent image (a measurement with the beam shutter closed) in case a CCD or CMOS detector is used, for each measurement duration in your measurement repertoire. Since the darkcurrent images often have a time-dependent and a time-independent component, it is necessary to measure the darkcurrent images for the same time as the actual sample measurement.

Thus, the user should determine for each sample:
\begin{itemize}
\item{$\square$ } The sample name (for logging)
\item{$\square$ } The sample filename
\item{$\square$ } The sample measurement duration 
\item{$\square$ } The sample transmission factor
\item{$\square$ } The sample thickness (thickness in the direction and location of the direct beam)
\item{$\square$ } The incident flux onto the sample
\item{$\square$ } Remarkable aspects regarding the sample
\item{$\square$ } The relevant background name (identifier for logging)
\item{$\square$ } The relevant background filename (all information collected for the sample (flux, transmission, etc.) should also be collected for the background measurement)
\item{$\square$ } When using image plates, the times of: the last erasure, the start of exposure, the end of exposure, the start of readout and the end of the readout procedure.
\end{itemize}

%\referencelist[achemsotwo] 
\bibliography{bibliography}

\end{document}